\documentclass{aastex} 
\def \be{\begin{equation}}
\def \ee{\end{equation}}
\def \bdm{\begin{eqnarray}}
\def \edm{\end{eqnarray}}
\begin{document}
\title{Simulations of Energetic Particles Interacting with Nonlinear Anisotropic Dynamical Turbulence}
\shorttitle{Particle Transport in Dynamical Turbulence}
\shortauthors{Heusen \& Shalchi}
\author{M. Heusen \& A. Shalchi}
\affil{Department of Physics and Astronomy, University of Manitoba, Winnipeg, Manitoba R3T 2N2, Canada}
\email{andreasm4@yahoo.com}
\begin{abstract}
We investigate test-particle diffusion in dynamical turbulence based on a numerical approach presented before.
For the turbulence we employ the nonlinear anisotropic dynamical turbulence model which takes into account
wave propagation effects as well as damping effects. We compute numerically diffusion coefficients of energetic
particles along and across the mean magnetic field. We focus on turbulence and particle parameters which
should be relevant for the solar system and compare our findings with different interplanetary observations.
We vary different parameters such as the dissipation range spectral index, the ratio of the turbulence
bendover scales, and the magnetic field strength in order to explore the relevance of the different parameters.
We show that the bendover scales as well as the magnetic field ratio have a strong influence on diffusion
coefficients whereas the influence of the dissipation range spectral index is weak. The best agreement with
solar wind observations can be found for equal bendover scales and a magnetic field ratio of $\delta B / B_0 = 0.75$.
\end{abstract}
\section{Introduction}
It is well-known that magnetic turbulence influences the motion of electrically charged energetic particles
such as cosmic rays. Turbulence in general has different properties such as the spectrum describing how the
magnetic energy is distributed among different length scales. Another fundamental aspect of turbulence is
spectral anisotropy describing how magnetic turbulence varies in different directions of space. Diffusion
of particles along the mean magnetic field, for instance, is controlled by gyro-resonant interactions (see, e.g.,
Schlickeiser 2002 and Shalchi 2009 for reviews). Therefore, the spectrum of turbulence at a certain scale
or wavenumber determines the diffusion coefficient of the energetic particles with a certain energy. It should
be emphasized, however, that nonlinear effects can be important for parallel diffusion and non-resonant interactions
can influence the diffusion parameter in certain parameter regimes (see Shalchi 2009 for a review). Spectral
anisotropy can also have an effect but this effect is weaker than originally thought (see Hussein et al. 2015).
For perpendicular diffusion, however, the details of the turbulence seem to be less important because
the perpendicular diffusion coefficient depends only on the so-called Kubo number and the parallel diffusion
coefficient (see Shalchi 2015). Due to the latter dependence, however, the perpendicular diffusion parameter
indirectly also depends on spectrum and spectral anisotropy.

Another important turbulence property is the dynamics describing the characteristic time scales over which
the turbulent magnetic field decorrelates. Different approaches have been proposed in the past to model the
turbulence dynamics. Some attempts are based on plasma wave propagation models in which the propagation effect
itself is taken into account as well as various damping effects (see again Schlickeiser 2002 for a review).
Or there is the important work of Bieber et al. (1994) in which simple models have been proposed to approximate
the temporal decorrelation of turbulence, namely the so-called {\it damping model of dynamical turbulence} and
the {\it random sweeping model}. In the recent years scientists achieved a more complete understanding
of the turbulence time scales. Therefore a more advanced model for the turbulence dynamics has been proposed
in Shalchi et al. (2006). This model is called the {\it Nonlinear Anisotropic Dynamical Turbulence (NADT) model}
and takes into account wave propagation effects as well as damping effects. It is the aim of this article
to simulate energetic particle motion in this type of turbulence and to explore the influence of different
turbulence parameters.

It was shown in different papers that dynamical turbulence effects can have a strong influence on the
transport of energetic particles. This concerns parallel diffusion (see, e.g., Bieber et al. 1994) but also
perpendicular diffusion (see, e.g., Shalchi et al. 2006). Such previous investigations were based on quasilinear
and nonlinear calculations. These days, however, one can also obtain diffusion parameters from test-particle
simulations. Previous work of this type was mostly done for magnetostatic turbulence (see, e.g., Giacalone \& Jokipii 1999,
Qin et al. 2002a, and Qin et al. 2002b) or undamped propagating plasma waves (see, e.g., Micha\l ek \& Ostrowski 1996
and Tautz \& Shalchi 2013). In Hussein \& Shalchi (2016) we have started to simulate test-particle transport in the
dynamical turbulence models used in Bieber et al. (1994), namely in the {\it damping model of dynamical turbulence}
and the {\it random sweeping model}. It was shown in Hussein \& Shalchi (2016) that for certain turbulence parameters
we can indeed reproduce different solar wind observations.

It is the purpose of the current paper to simulate particle transport in the more realistic NADT model and to
compute the parallel mean free path $\lambda_{\parallel}$, the perpendicular mean free path $\lambda_{\perp}$,
and the ratio of the two mean free paths $\lambda_{\perp} / \lambda_{\parallel}$. As in Hussein \& Shalchi (2016)
our findings are compared with the Palmer (1982) consensus range, observations of Jovian electrons (see Chenette et al. 1977),
and Ulysses measurements of Galactic protons (see Burger et al. 2000). We also explore how the different turbulence
parameters influence the different diffusion parameters.

The reminder of the paper is organized as follows. In Section 2 we explain the physics of turbulence in general but we focus
on the NADT model used in the current paper. The methodology which is used to perform particle transport simulations in
dynamical turbulence is explained in Section 3. In Section 4 we show our numerical results obtained for parallel and perpendicular
diffusion coefficients and we compare them with different solar wind observations. In Section 5 we conclude and summarize.
\section{Dynamical Turbulence}
\subsection{Description of Magnetic Turbulence}
In the analytical description of turbulence, the fundamental quantity is the magnetic correlation tensor in the wave vector space.
The components of the latter tensor are defined via
\be
P_{mn} \left( \vec{k}, t \right) = \left< \delta B_m \left( \vec{k}, t \right) \delta B_n^* \left( \vec{k}, 0 \right) \right>
\ee
where we have used the ensemble average operator $\langle\dots\rangle$. A standard assumption in the theory of dynamical turbulence
is that all tensor components obey the same temporal behavior and, therefore, they can be written as
\be
P_{mn} \left( \vec{k}, t \right) = P_{mn} \left( \vec{k} \right) \Gamma \left( \vec{k}, t \right).
\label{tensorwithgamma}
\ee
Here we have used the magnetostatic tensor components $P_{mn} ( \vec{k} )$ and the {\it dynamical correlation function} $\Gamma ( \vec{k}, t )$.
In the current paper we employ the NADT model in order to approximate the function $\Gamma ( \vec{k}, t )$. Before we discuss this model in detail,
we focus on the static tensor components.
\subsection{Two-Component Turbulence}
The slab/2D composite model is widely used in the transport theory of energetic particles (see, e.g., Bieber et al. 1994 and
Bieber et al. 1996). In the current paper we employ this model, which is also known as two-component model, as it was already
done in Hussein \& Shalchi (2016). This type of turbulence description is supported by observations in the solar wind (see, e.g.,
Matthaeus et al. 1990, Osman and Horbury 2009a, Osman and Horbury 2009b, Turner et al. 2012), turbulence simulations (see, e.g.,
Oughton et al. 1994, Matthaeus et al. 1996, Shaikh and Zank 2007) as well as analytical treatments of turbulence (see, e.g.,
Zank and Matthaeus 1993). More details concerning the used model can be found in the aforementioned articles or in the corresponding
diffusion theory papers (see, e.g., Hussein et al. (2015) and Hussein \& Shalchi (2016)).

Within the two-component approximation, the components of the static correlation tensor are written as
\be
P_{mn} = P_{mn}^{slab} + P_{mn}^{2D}
\ee
where we have used the components of the slab tensor
\be
P_{mn}^{slab} (\vec{k}) = g^{slab}(k_{\parallel}) \frac{\delta (k_{\perp})}{k_{\perp}} \delta_{mn},
\label{Plmslab}
\ee
and the components of the two-dimensional tensor
\be
P_{mn}^{2D} (\vec{k}) = g^{2D} (k_{\perp}) \frac{\delta (k_{\parallel})}{k_{\perp}} \left( \delta_{mn} - \frac{k_m k_n}{k_{\perp}^2} \right),
\label{Plm2D}
\ee
with $m,n=x,y$. Furthermore, we have $P_{mz}=P_{zn}=P_{zz}=0$ in both cases due to $\delta B_z = 0$. For the two-dimensional modes,
the latter assumption is motivated by the fact that in the solar wind the power in parallel fluctuations is small in the inertial
range (see Belcher \& Davis 1971). For the slab modes $\delta B_z = 0$ is a consequence of the solenoidal constraint $\nabla \cdot \vec{B} = 0$.

In Eqs. (\ref{Plmslab}) and (\ref{Plm2D}) we have used the slab spectrum $g^{slab}(k_{\parallel})$ as well as the two-dimensional (2D) spectrum
$g^{2D} (k_{\perp})$, respectively. For the former spectrum we employ the form
\bdm
g^{slab} (k_{\parallel}) & = & \frac{C(s)}{2 \pi} l_{slab} \delta B_{slab}^2 \nonumber\\
& \times & \left\{
\begin{array}{ccc}
(1 + k_{\parallel}^2 l_{slab}^2)^{-s/2} & \textnormal{if} & k_{\parallel} \leq k_{d} \\
(1+k_{d}^2 l_{slab}^2)^{-s/2} (k_{d} / k_{\parallel})^{p} & \textnormal{if} & k_{\parallel} \geq k_{d}
\end{array}
\right.
\label{gslab}
\edm
as proposed in Bieber et al. (1994). Here we have used the slab bendover scale $l_{slab}$, the dissipation wavenumber $k_d$, the inertial range
spectral index $s$, and the dissipation range spectral index $p$. Furthermore, we have employed the normalization function
\be
C(s) = \frac{\Gamma \left( \frac{s}{2} \right)}{2 \sqrt{\pi} \Gamma \left( \frac{s-1}{2} \right)}
\label{normalC}
\ee
with the Gamma function $\Gamma (z)$. The spectrum is correctly normalized as long as $s > 1$.

\begin{figure}
\centering
\includegraphics[scale=0.5]{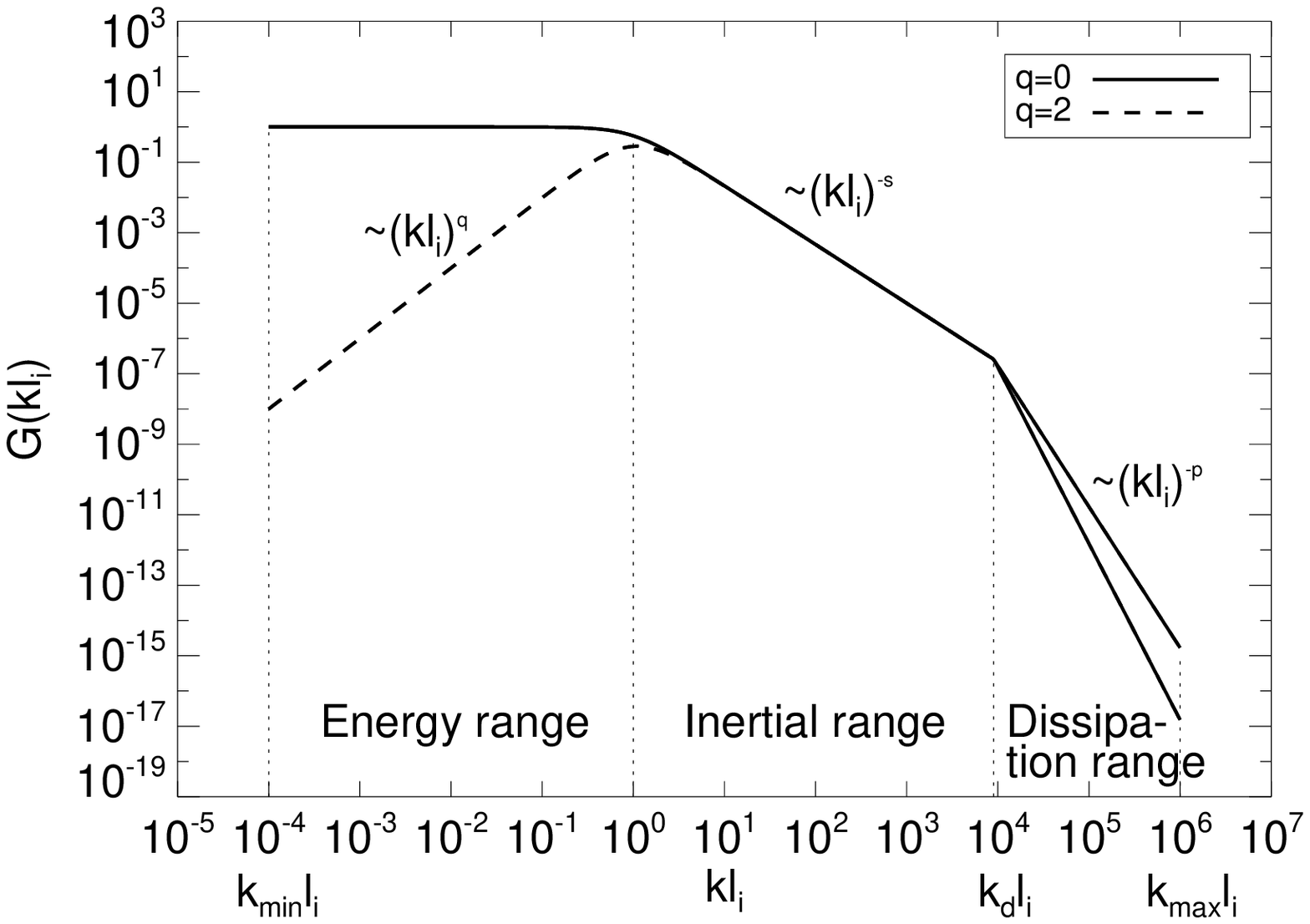}
\caption{The spectra used in the current paper for the slab modes ($q=0$) and the two-dimensional modes ($q=2$).}
\label{thespectrum}
\end{figure}

For the two-dimensional spectrum we use an extension of the model proposed by Bieber et al. (1994). By combining the spectrum used in the
latter paper with the ideas discussed in Matthaeus et al. (2007) and Shalchi \& Weinhorst (2009), we propose the form
\bdm
g^{2D} (k_{\perp}) & = & \frac{2 D(s,q)}{\pi} l_{2D} \delta B_{2D}^2 \nonumber\\
& \times & \left\{
\begin{array}{ccc}
\frac{(k_{\perp} l_{2D})^{q}}{(1 + k_{\perp}^2 l_{2D}^2)^{(s+q)/2}} & \textnormal{if} & k_{\perp} \leq k_{d} \\
\frac{(k_{d} l_{2D})^{q}}{(1+k_{d}^2 l_{2D}^2)^{(s+q)/2}} \left( \frac{k_{d}}{k_{\perp}} \right)^{p} & \textnormal{if} & k_{\perp} \geq k_{d}.
\end{array}
\right.
\label{g2d}
\edm 
The only parameter which is different compared to the slab spectrum, is the energy range spectral index $q$ controlling the spectral shape
at large turbulence scales. Furthermore, we have used the extended normalization function
\be
D(s, q) = \frac{\Gamma \left( \frac{s+q}{2} \right)}{2 \Gamma \left( \frac{s-1}{2} \right) \Gamma \left( \frac{q+1}{2} \right)}
\label{normalD}
\ee
with $s > 1$ and $q > -1$. Eqs. (\ref{normalC}) and (\ref{normalD}) are linked via $C(s) = D(s, q=0)$. In Tables \ref{simvalues} and \ref{runs}
we list the values we have used in our simulations for the different turbulence and particle parameters. In Fig. \ref{thespectrum} we visualize
the used spectra for slab and two-dimensional modes, respectively.

A spacial aspect of the two-component model used here is that we assume that there are no fluctuations parallel to the mean field $\delta B_z = 0$.
More recent observations (see, e.g., Alexandrova et al. 2008) and numerical simulations (see, e.g., Howes et al. 2008) show an increased level of
magnetic compressibility at small scales. In Hussein et al. (2015) the influence of different magnetostatic turbulence models on the parallel and
perpendicular diffusion coefficients was explore numerically. No strong influence was found indicating that a non-vanishing turbulent field in the
parallel direction is less important. However, the latter statement is not true for very strong turbulence in which the turbulent field is much
stronger than the mean field (see Hussein \& Shalchi 2014). In such cases we find isotropic diffusion meaning that the parallel and perpendicular
diffusion coefficients are equal.

Furthermore, the turbulence model used in the current paper is axi-symmetric with respect to the mean magnetic field. Observations (see, e.g, Saur \& Bieber
1999 and Narita et al. 2010) and numerical simulations (see, e.g., Dong et al. 2014) have shown that solar wind turbulent spectra are not axi-symmetric.
If deviations from axi-symmetry are take into account, the whole diffusion tensor needs to be computed (see, e.g., Weinhorst et al. 2008). It will
be subject of future work to present a detailed numerical investigation of test-particle transport in turbulent systems without axi-symmetry.

\begin{table}
\caption{The parameter values used for our test-particle simulations. The values should be appropriate in the interplanetary space
at $1$ AU heliocentric distance (see, e.g., King 1989).}
\renewcommand{\arraystretch}{1.5}
\begin{tabular}{lll}
\hline
Parameter							& Symbol				& Value								\\ 
\hline
2D energy range spectral index		& $ q $ 				& $ 2 $ 							\\
2D inertial range spectral index	& $ s^{2D} $			& $ 5/3	$ 							\\
Alfv\'en speed						& $ v_A $ 				& $ 33.5 $ km/s  					\\
Slab bendover scale					& $ l_{slab} $ 			& $ 0.030$ AU	 					\\
Slab dissipation wavenumber			& $ k_{d}^{slab} $		& $ 3 \times 10^{5}$ 1/AU		 	\\
Mean magnetic field					& $ B_0	$ 				& $ 4.12$ nT						\\
Slab fraction						& $ \delta B_{slab}^2 $ & $ 0.2 \; \delta B^2 $ 			\\
2D fraction							& $ \delta B_{2D}^2 $ 	& $ 0.8 \; \delta B^2 $ 			\\
\hline
\end{tabular}
\medskip
\label{simvalues}
\end{table}
\begin{table}[t]
\caption{The different runs performed in the current paper and the values used for the relative turbulence strength $\delta B / B_0$, the slab
inertial range spectral index $s^{slab}$, the dissipation range spectral index $p$, the ratio of the two bendover scales $l_{2D}/l_{slab}$, and
the two-dimensional dissipation wavenumber $k_d^{2D}/k_d^{slab}$.}
\begin{center}
\renewcommand{\arraystretch}{1.5}
\begin{tabular}{ccccccc}
\hline
Section	& $\delta B / B_0$	& $s^{slab}$	& $p$				& $l_{2D}/l_{slab}$		& $k_d^{2D}/k_d^{slab}$	& Figures										\\
\hline
1		& $ 1 $				& $5/3$			& $3$				& $1$					& $1$					& \ref{paracompdb1}-\ref{ratiocompdb1}			\\
2		& $ 0.5 $			& $5/3$			& $3$				& $1$					& $1$					& \ref{paracompdb05}-\ref{ratiocompdb05}		\\
3		& $ 0.75 $			& $5/3$			& $3$				& $1$					& $1$					& \ref{paracompdb075}-\ref{ratiocompdb075}		\\
4		& $ 0.5 $			& $5/3$			& $3$, $4$, $5$		& $1$					& $1$					& \ref{paracompdb05ppp}-\ref{ratiocompdb05ppp}	\\
5		& $ 0.5 $			& $5/3$			& $3$				& $0.1$					& $1$					& \ref{paracompdb05l2d}-\ref{ratiocompdb05l2d}	\\
6		& $ 0.75 $			& $2$			& $3$				& $1$					& $10$					& \ref{prll_diss}-\ref{perp_prll_diss}		\\
7		& $ 0.75 $			& $2$			& $3$				& $1$					& $10$					& \ref{prll_index}-\ref{perp_prll_index}		\\
\hline
\end{tabular}
\end{center}
\label{runs}
\end{table}
\subsection{The Nonlinear Anisotropic Dynamical Turbulence Model}
In order to model dynamical turbulence, one has to specify the dynamical correlation function $\Gamma (\vec{k},t)$ in
Eq. (\ref{tensorwithgamma}). In recent years there has been a more complete understanding of the time scales of turbulence
(see, e.g., Matthaeus et al. 1990, Tu \& Marsch 1993, Zhou et al. 2004, and Oughton et al. 2006). Based on this improved
understanding, Shalchi et al. (2006) have developed the NADT model for the function $\Gamma (\vec{k},t)$. Within the latter
model, we have different dynamical correlation functions for slab and two-dimensional modes, respectively.

For the corresponding function of the slab modes we have according to Shalchi et al. (2006)
\be
\Gamma^{slab} (k_{\parallel},t) = e^{i \omega_p t - \beta t}
\label{c2e8}
\ee
where we have used the constant
\be
\beta = \sqrt{2} \alpha {v_A \over l_{2D}} {\delta B_{2D} \over B_0}
\label{defbeta}
\ee
and the plasma wave dispersion relation of shear Alfv\'en waves 
\be
\omega_p = j v_A k_{\parallel}.
\label{c2e10}
\ee
Obviously one finds an oscillating factor in Eq. (\ref{c2e8}) describing wave propagation effects. The parameter $j$
used in Eq. (\ref{c2e10}) indicates the wave propagation direction. Here $j=+1$ is used for forward and $j=-1$ for backward
to the ambient magnetic field propagating waves. One would expect that closer to the sun the most waves
should propagate forward and far away from the sun the wave intensities should be equal for both directions
(see, e.g., Bavassano 2003 for more details). In the current paper we are interested in turbulence parameters
at $1$ AU heliocentric distance and, thus, we assume that all waves propagate forward and we set $j=+1$. The exponential
factor in Eq. (\ref{c2e8}) contains the decorrelation time scale $\tau = 1/\beta$ where $\beta$ is given by Eq. (\ref{defbeta}).
The slab component in our model is assumed to experience resonant nonlinear triad interactions with the low-frequency
two-dimensional component. Therefore, the time $\tau$ is given by the global two-dimensional nonlinear timescale.
The parameter $\alpha$ in Eq. (\ref{defbeta}) is a constant of order one related to the so-called {\it Karman-Taylor
constant} and $v_A$ is the Alfv\'en speed. In the current paper we set $\alpha = 1$ for simplicity.

For the two-dimensional modes we have according to Shalchi et al. (2006)
\be
\Gamma^{2D} (k_{\perp},t) = e^{- \gamma t}
\label{c2e11}
\ee
where we have used
\bdm
\gamma = \gamma (k_{\perp}) = \beta
\left\{
\begin{array}{ccc}
1 & \quad \textnormal{for} \quad & k_{\perp} l_{2D} \leq 1 \\
(k_{\perp} l_{2D})^{2/3} & \textnormal{for} & k_{\perp} l_{2D} \geq 1
\end{array}
\right.
\label{c2e12}
\edm
with the constant $\beta$ defined already in Eq. (\ref{defbeta}). Obviously no oscillatory factor appears in Eq. (\ref{c2e11}).
For small perpendicular wavenumbers $k_{\perp}$, we estimate the correlation time as above for the slab modes. For large $k_{\perp}$,
however, the decorrelation time is estimated by using a steady inertial range $k_{\perp}^{-5/3}$ approximation.

In analytical treatments of the transport, one can directly use the models described here. As pointed out in
Hussein \& Shalchi (2016), this is not the case in test-particle simulations where a Fourier transformation
has to be employed for the dynamical correlation function. We define
\be
\chi (\vec{k}, \omega) := \frac{1}{\pi} \Re \int_{0}^{\infty} d t \; \Gamma (\vec{k},t) e^{- i \omega t}.
\ee
Using $\chi (\vec{k}, \omega)$ instead of $\Gamma (\vec{k},t)$ means that we describe the turbulence in a four-dimensional
Fourier space with the coordinates $\vec{k}$ and $\omega$.

In the NADT model, the dynamical correlation function for the slab modes is given by Eq. (\ref{c2e8}). Therefore,
we find
\be
\chi^{slab} (\vec{k}, \omega) := \frac{1}{\pi} \frac{\beta}{\beta^2 + \left( \omega - \omega_p \right)^2}
\label{chislab}
\ee
where $\omega_p = \omega_p (\vec{k})$ is given by Eq. (\ref{c2e10}). For the two-dimensional modes, the
dynamical correlation function is given by Eq. (\ref{c2e11}) and, thus
\be
\chi^{2D} (\vec{k}, \omega) := \frac{1}{\pi} \frac{\gamma}{\gamma^2 + \omega^2}
\label{chi2d}
\ee
where $\gamma = \gamma (\vec{k})$ is given by Eq. (\ref{c2e12}).

In the next section we explain our numerical approach and in Sect. 4 we show the results for the turbulence model
described here.
\section{Methodology}
We simulate particle transport in dynamical turbulence based on the method described in Hussein \& Shalchi (2016).
The first step is the creation of turbulence by using the formula
\be
\delta \vec{B} (\vec{x},t)=\sqrt{2} \; \delta B \sum_{m=1}^{M} \sum_{n=1}^{N} A \left( k_m, \omega_n \right)
\vec{\xi}_m e^{i \left( \vec{k}_m \cdot \vec{x} + \omega_n t + \beta_{mn} \right)}
\label{fieldscode}
\ee
with the random phase $\beta_{mn}$. The used method can be seen as an extension of previous simulations performed
for either magnetostatic turbulence or undamped propagating plasma waves (see, e.g., Micha\l ek \& Ostrowski 1996,
Giacalone \& Jokipii 1999, Tautz 2010, and Hussein et al. 2015). In the following we describe the parameters and
functions used in Eq. (\ref{fieldscode}).

We create two-component turbulence by employing Eq. (\ref{fieldscode}) for slab and two-dimensional modes, respectively
and then we add the two obtained magnetic field vectors. For the slab modes and the two-dimensional modes we use the
same polarization vector $\vec{\xi}_m$, namely 
\be
\vec{\xi}_m = \left( -sin\phi_m, cos\phi_m, 0 \right)
\label{defxi}
\ee
where $\phi_m$ is a random angle.

All quantities used in the code are normalized with respect to the slab bendover scale $l_{slab}$. This means, for instance, that $k_m$
used above corresponds to the physical quantity $k_{\parallel} l_{slab}$ or $k_{\perp} l_{slab}$ and $z$ stands for $z / l_{slab}$.
The frequency $\omega$ is normalized with respect to the unperturbed gyro frequency of the particle $\Omega=(q B_0)/(m c \gamma)$
meaning that $\omega_n=\omega/\Omega$. Here we have used the electric charge of the particle $q$, the rest mass $m$, the speed of
light $c$, and the Lorentz factor $\gamma$.

In Eq. (\ref{fieldscode}) we have also used $\vec{k}_m = k_m \hat{k}_m$ with the random wave unit vector
\be
\hat{k}_m=
\left(
\begin{array}{c}
\sqrt{1-\eta_m^2}\cos \phi_m \\
\sqrt{1-\eta_m^2}\sin \phi_m \\
\eta_m
\end{array}
\right).
\ee
The random angle $\phi_m$ was already used in Eq. (\ref{defxi}). What the value of $\eta_m$ is depends on the simulated turbulence model.
For the slab modes we have $\eta_m = 1$ and for two-dimensional modes $\eta_m = 0$. In Eq. (\ref{fieldscode}) we have also used the amplitude
function
\be
A^2(\omega_n,k_m)=\frac{G \left( k_m, \omega_n \right) \Delta k_m \Delta \omega_n}{\sum_{\mu=1}^M \sum_{\nu=1}^N G( k_{\mu}, \omega_{\nu}) \Delta k_{\mu} \Delta \omega_{\nu}}
\ee
where $G(k_{\mu}, \omega_{\nu})$ represents the space-time spectrum
\be
G \left( k_m, \omega_n \right) = G(k_m) \chi (k_m, \omega_n).
\ee
Eqs. (\ref{chislab}) and (\ref{chi2d}) show the functions $\chi (k_m, \omega_n)$ for the NADT model. The function $G(k_m)$ is the usual
spectrum as used in simulations of magnetostatic turbulence (see, e.g., Hussein et al. 2015). In the current paper we employ Eq. (\ref{gslab})
for the slab modes and Eq. (\ref{g2d}) for the two-dimensional modes.

In the used model for $\chi (k_m, \omega_n)$, one finds the Alfv\'en speed $v_A$ which can be normalized with respect to the particle
speed $v$ so that
\be
\frac{v_A}{v} = \frac{v_A}{c} \frac{\sqrt{R_0^2 + R^2}}{R}.
\ee
Here we have used the parameter
\bdm
R_0 = \frac{1}{l_{slab} B_0}
\left\{
\begin{array}{ccc}
0.511 \textnormal{MV} \quad & \textnormal{for} \quad & \textnormal{electrons} \\
938 \textnormal{MV} \quad & \textnormal{for} \quad & \textnormal{protons},
\end{array}
\right.
\label{defR0}
\edm
and the dimensionless rigidity defined via $R = R_L / l_{slab}$ where $R_L = v / \Omega$ is the unperturbed Larmor radius.
For $l_{slab} = 0.03$AU and $B_0 = 4.12$nT this gives $R_0=9.2\times10^{-5}$ for electrons and $R_0=0.169$ for protons.
All other parameter values used in our simulations are listed in Tables \ref{simvalues} and \ref{runs} .

In our test-particle simulations in dynamical turbulence we have to deal with the same problems one has to deal with in
simulations of static turbulence (see again Hussein et al. (2015) for more details). For dynamical turbulence, however, there
are a few additional concerns. We need a certain number of grid points in space and time. For most of our runs we have set $N=M=256$
in Eq. (\ref{fieldscode}). For lower rigidities we had to set $N=M=64$ to avoid too long computation times. In all runs we have
computed running diffusion coefficients for times up to at least $\Omega t = 10^4$ (here $\Omega$ denotes again the unperturbed
gyro frequency) to ensure that we are in the stable regime.

Furthermore, as noted in Hussein \& Shalchi (2016), the value of the minimum frequency $\omega_{min}$ has a strong influence on
the obtained parallel and perpendicular mean free paths. This influence was noticed for both protons and electrons but was much
stronger for electrons. To avoid this problem, we performed our simulations for small enough values of $\omega_{min}$.

Following the ideas presented in Tautz (2010), we also compute the errors of the different mean free paths. The latter
author noted that using the standard deviation as a mean of estimating the error is inappropriate as the mean square
displacement calculated in the Monte Carlo code is the variance of the distribution function for the diffusion equation
itself. In addition, test particles interacting with turbulent magnetic fields scatter in a random manner leading to a
huge variance in their square deviation. Hence one has to come up with a method that takes into account the averaging
processes used over the number of turbulence manifestations, $N_{T}$, for each of which a fixed number of test particles
were simulated in space and time resulting a diffusion coefficient. The mean error is then defined to be the deviation
of the different mean free paths $\lambda_n$ from the final averaged mean free path $\lambda_{f}$. Mathematically this
reads
\bdm
\sigma_{\lambda}^2 & = & \frac{1}{N_{T}-1} \nonumber\\
& \times & \Bigg \{ \sum_{n=0}^{N_{T}}(\lambda_n-\lambda_{f})^2 -\frac{1}{N_{T}}\bigg[\sum_{n=0}^{N_{T}}(\lambda_n-\lambda_{f})\bigg]^2 \Bigg \}.
\label{error}
\edm 
Using Eq. (\ref{error}), both the error in parallel and perpendicular mean free paths where calculated, $\Delta \lambda_{\parallel}$ and
$\Delta \lambda_{\perp}$ respectively. To calculate the error in the ratio of the two mean free paths, $\lambda_{\perp}/\lambda_{\parallel}$,
we use the rule of error combination 
\be
\Delta \bigg(\frac{\lambda_{\perp}}{\lambda_{\parallel}}\bigg )= \bigg ( \frac{\Delta \lambda_{\perp}}{\lambda_{\perp}} + \frac{\Delta \lambda_{\parallel}}{\lambda_{\parallel}} \bigg ) \frac{\lambda_{\perp}}{\lambda_{\parallel}}.
\ee
In most plots shown in Sect. 4 we have included the error bars based on the method presented here.
\section{Results}
In Tables \ref{simvalues} and \ref{runs} we show the different parameters used in our simulation runs. In the following we vary
the magnetic field ratio $\delta B / B_0$, the dissipation range spectral index $p$, the ratio of the bendover scales $l_{2D} / l_{slab}$,
the inertial range spectral index of the slab modes $s^{slab}$, as well as the dissipation wavenumber of the two-dimensional
modes $k_d^{2D}$.
\subsection{Slab/2D Turbulence with $\delta B / B_0 = 1.0$}
\begin{figure}
\centering
\includegraphics[scale=0.5]{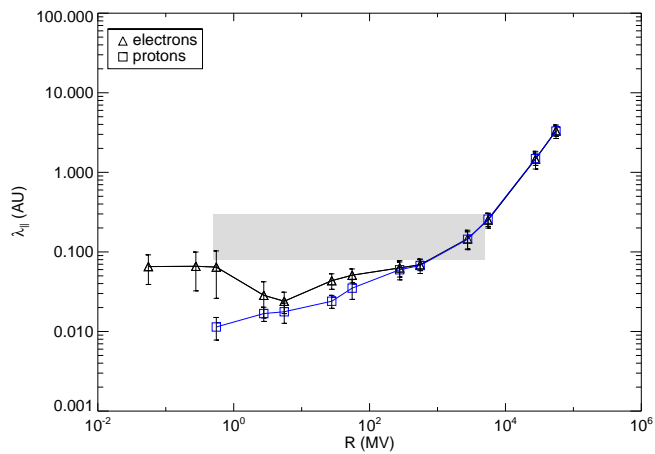}
\caption{The parallel mean free path versus magnetic rigidity for two-component turbulence, the NADT model,
and $\delta B / B_0 = 1.0$. The shaded band represents the Palmer (1982) consensus range.}
\label{paracompdb1}
\end{figure}

\begin{figure}
\centering
\includegraphics[scale=0.5]{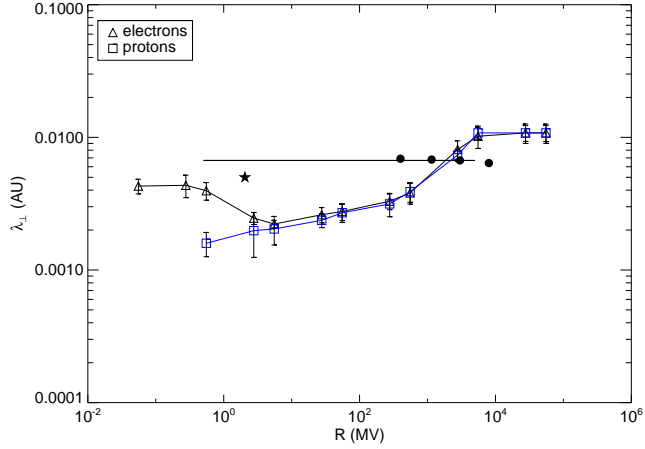}
\caption{The perpendicular mean free path versus magnetic rigidity for two-component turbulence, the NADT model,
and $\delta B / B_0 = 1.0$. For comparison we show observations of Jovian electrons (Chenette et al. 1977, star), Ulysses measurements 
of Galactic protons (Burger et al. 2000, dots), and the Palmer (1982) value (horizontal line).}
\label{perpcompdb1}
\end{figure}

\begin{figure}
\centering
\includegraphics[scale=0.5]{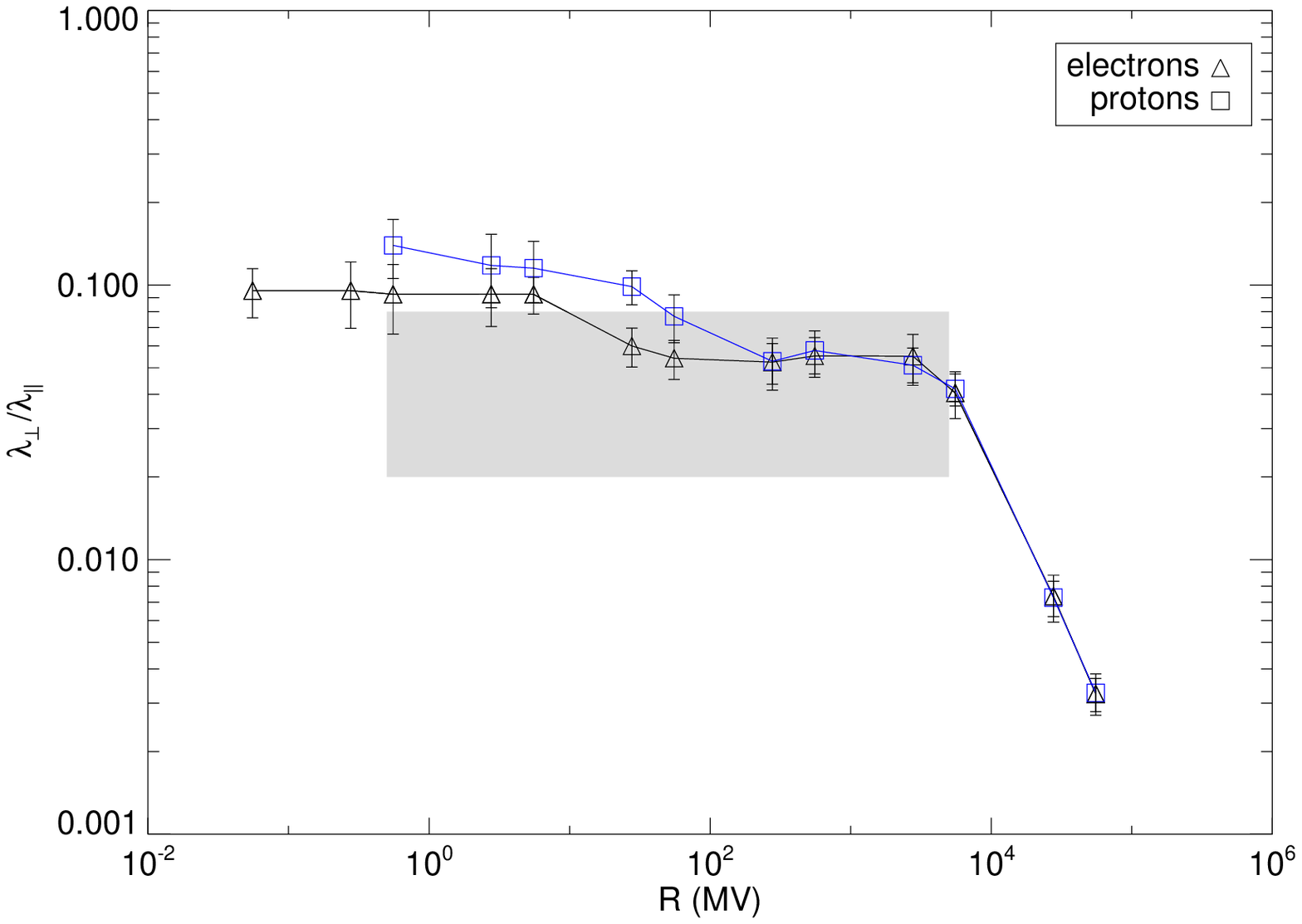}
\caption{The ratio of perpendicular and parallel mean free paths versus magnetic rigidity for two-component turbulence, the NADT model,
and $\delta B / B_0 = 1.0$. The shaded band represents the Palmer (1982) consensus range.}
\label{ratiocompdb1}
\end{figure}

Often one assumes that the ratio of turbulent and mean magnetic field is $\delta B / B_0 = 1.0$ (see, e.g., Bieber et al. 1994
and Bieber et al. 1996). Furthermore, we assume equal turbulence bendover scales $l_{2D} = l_{slab}$ and set the dissipation
range spectral index to $p=3$. We vary the particle rigidity from usually a few percent megavolt up to about $50$ gigavolt.
We compute the parallel mean free path, the perpendicular mean free path, as well as the ratio of the two diffusion parameters
$\lambda_{\perp} / \lambda_{\parallel}$. Our numerical findings are visualized in Figs. \ref{paracompdb1}, \ref{perpcompdb1},
and \ref{ratiocompdb1}. All results are compared with different measurements performed in the solar system.

Qualitatively, our results are similar compared to the simulations presented in Hussein \& Shalchi (2016) which were obtained
for the {\it damping model of dynamical turbulence} and the {\it random sweeping model}. As in previous work we conclude that
the obtained parallel mean free paths are too small compared to the Palmer (1982) consensus range. Therefore, we change
different parameters in our test-particle code to explore their influence on the different diffusion parameters. This is done
in the following paragraphs.

\subsection{Slab/2D Turbulence with $\delta B / B_0 = 0.5$}
\begin{figure}
\centering
\includegraphics[scale=0.5]{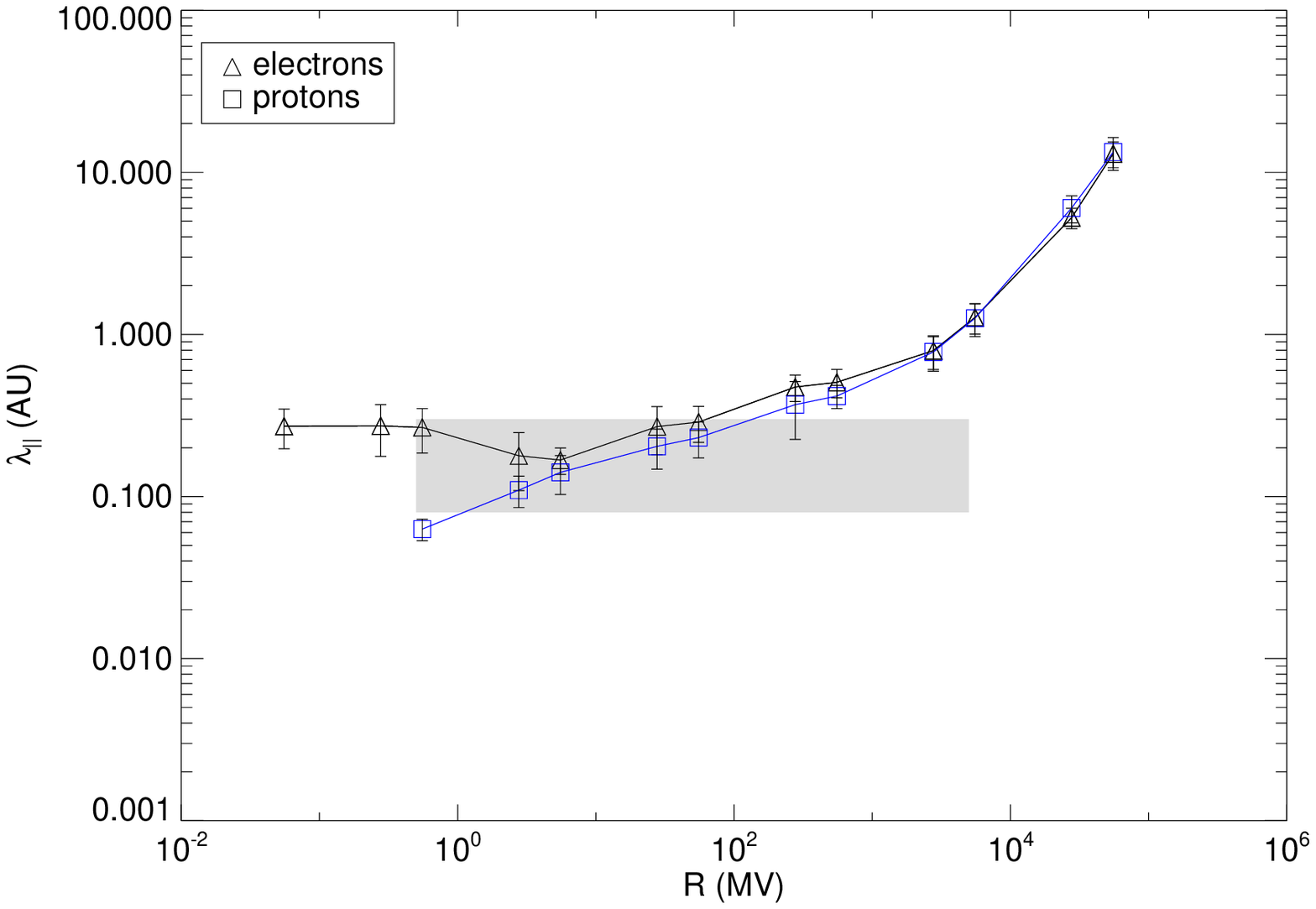}
\caption{Caption is exactly as in Fig. \ref{paracompdb1} but results were obtained for $\delta B / B_0 = 0.5$.}
\label{paracompdb05}
\end{figure}

\begin{figure}
\centering
\includegraphics[scale=0.5]{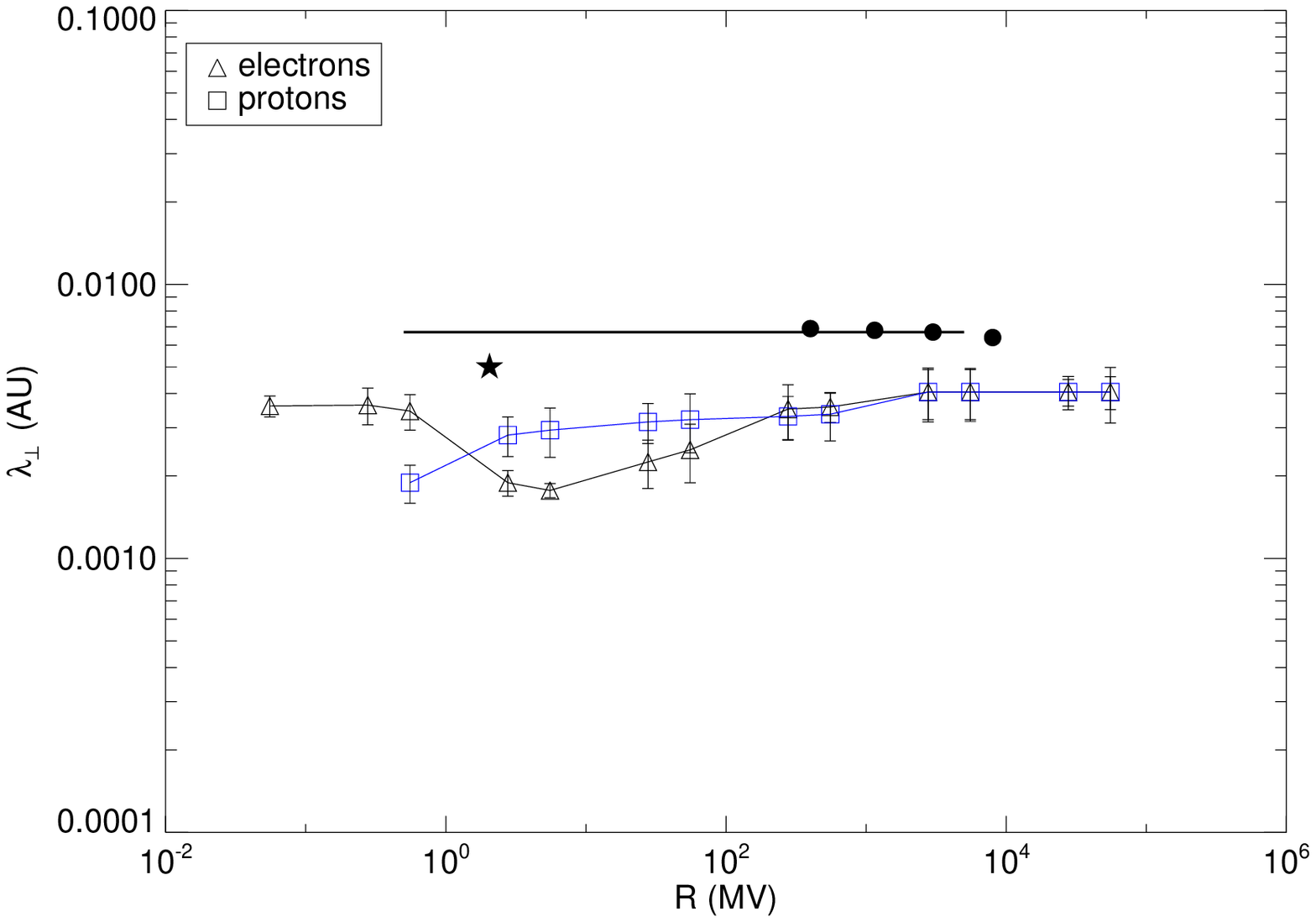}
\caption{Caption is exactly as in Fig. \ref{perpcompdb1} but results were obtained for $\delta B / B_0 = 0.5$.}
\label{perpcompdb05}
\end{figure}

\begin{figure}
\centering
\includegraphics[scale=0.5]{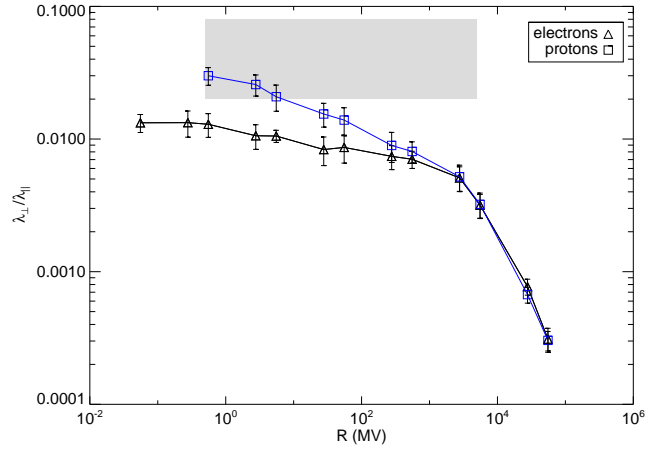}
\caption{Caption is exactly as in Fig. \ref{ratiocompdb1} but results were obtained for $\delta B / B_0 = 0.5$.}
\label{ratiocompdb05}
\end{figure}

In Hussein \& Shalchi (2016) it was shown that the simulated parallel mean free path is too small if the magnetic field ratio
is assumed to be $\delta B / B_0 = 1$. Therefore, the latter ratio was changed to $\delta B / B_0 = 0.5$ as suggested in Ruffolo et al. (2012).
In the current paragraph we do the same in the context of the NADT model and we show our findings for the different diffusion parameters
in Figs. \ref{paracompdb05}, \ref{perpcompdb05}, and \ref{ratiocompdb05}.

As expected we find an increased parallel mean free path but a smaller perpendicular mean free path. The former transport coefficient
goes directly through the Palmer (1982) consensus range confirming that we can indeed reproduce solar wind observations of energetic
particles numerically. The perpendicular diffusion coefficients, however, are now too small. The same applies for the ratio of the two
diffusion parameters $\lambda_{\perp} / \lambda_{\parallel}$.

\subsection{Slab/2D Turbulence with $\delta B / B_0 = 0.75$}
\begin{figure}
\centering
\includegraphics[scale=0.5]{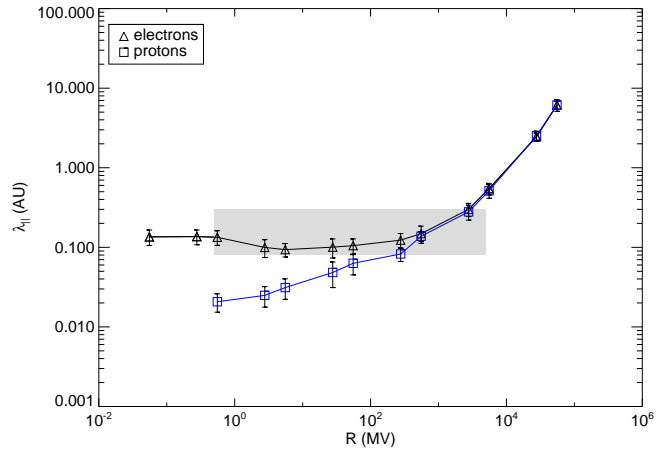}
\caption{Caption is exactly as in Fig. \ref{paracompdb1} but results were obtained for $\delta B / B_0 = 0.75$.}
\label{paracompdb075}
\end{figure}

\begin{figure}
\centering
\includegraphics[scale=0.5]{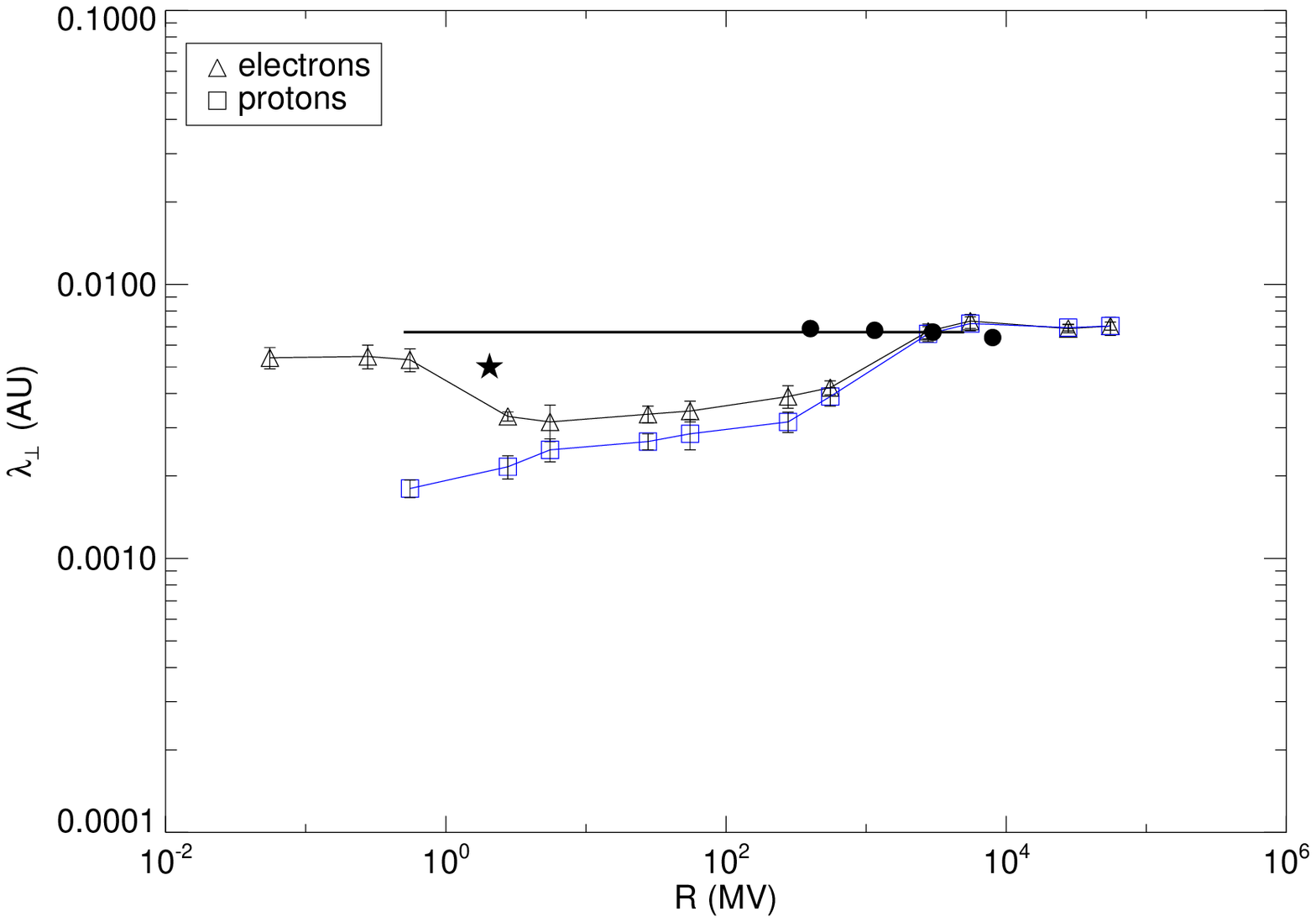}
\caption{Caption is exactly as in Fig. \ref{perpcompdb1} but results were obtained for $\delta B / B_0 = 0.75$.}
\label{perpcompdb075}
\end{figure}

\begin{figure}
\centering
\includegraphics[scale=0.5]{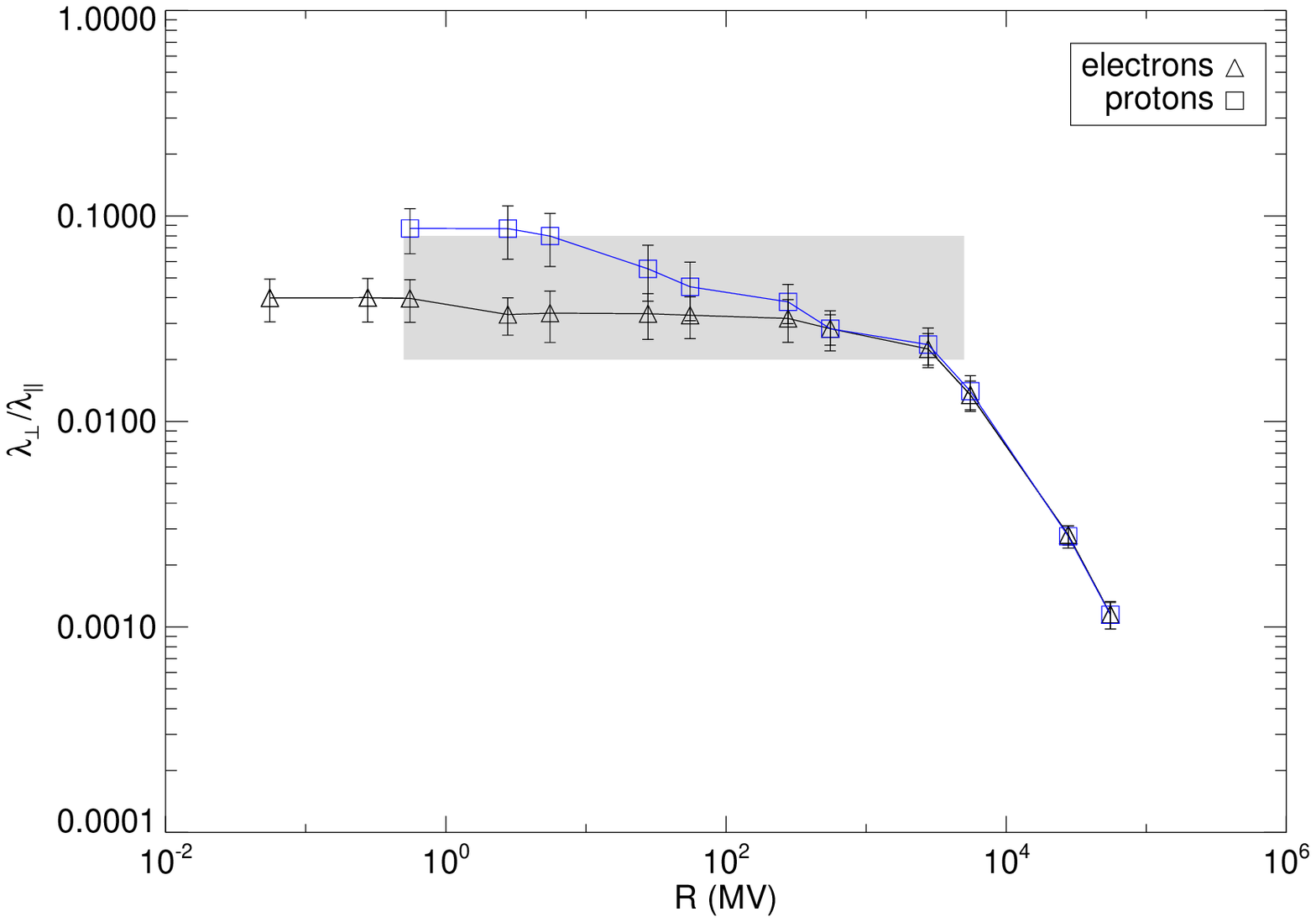}
\caption{Caption is exactly as in Fig. \ref{ratiocompdb1} but results were obtained for $\delta B / B_0 = 0.75$.}
\label{ratiocompdb075}
\end{figure}

Above we have performed the simulations for the magnetic field ratios $\delta B / B_0 = 1$ and $\delta B / B_0 = 0.5$. According to
Fig. \ref{paracompdb1} the parallel mean free path is too short for $\delta B / B_0 = 1$. For a reduced magnetic field ratio of
$\delta B / B_0 = 0.5$ the parallel mean free path is much larger but is still within the Palmer (1982) consensus range (see Fig. \ref{paracompdb05}).
In the current paragraph we show the simulations performed for an intermediate turbulence level of $\delta B / B_0 = 0.75$.
The obtained diffusion parameters are visualized in Figs. \ref{paracompdb075}, \ref{perpcompdb075}, and \ref{ratiocompdb075}.

As expected, the parallel mean free path for electrons is now perfectly inside the box representing the solar wind observations.
The perpendicular mean free path as well as the ratio of the two diffusion coefficients is close to the different observations as well.
Obviously, the magnetic field ratio is a critical parameter controlling both spatial diffusion coefficients. This is exactly what one
expects and what is also predicted by analytical investigations of the transport (see, e.g., Shalchi 2009 and Shalchi 2015).
For $\delta B / B_0 = 0.75$ we find the best agreement between simulations and observations.

\subsection{Influence of the Dissipation Range Spectral Index}
Above, as well as in Hussein \& Shalchi (2016), the dissipation range spectral index was set to $p=3$. This is a numerical
value which is close to solar wind observations of magnetic turbulence (see, e.g., Denskat \& Neubauer 1982). It is expected
that the smallest scales of turbulence, corresponding to the dissipation range, influence the parallel mean free path at
low rigidities due to the gyroresonant interactions between particles and turbulence.

\begin{figure}
\centering
\includegraphics[scale=0.5]{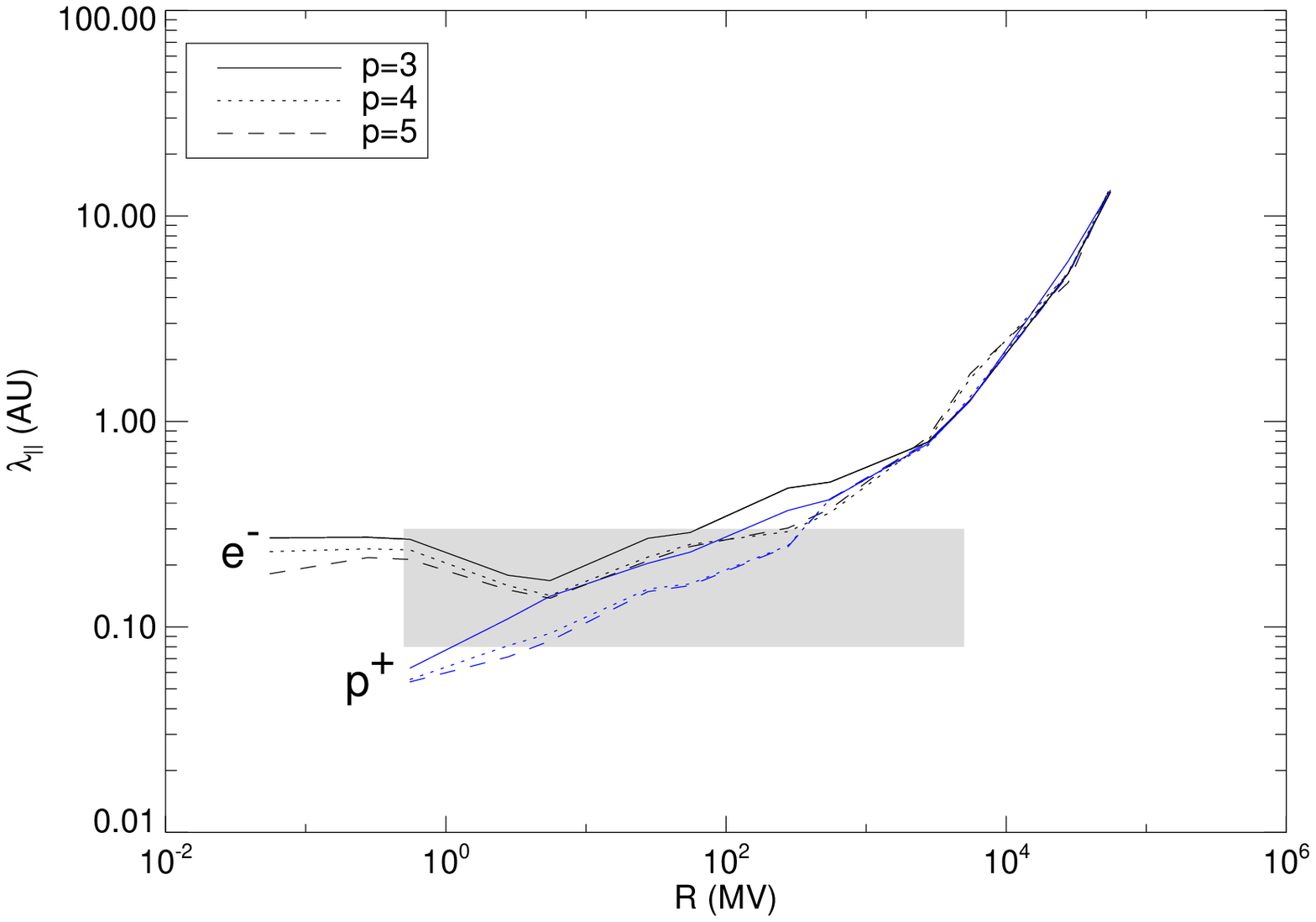}
\caption{The parallel mean free path versus magnetic rigidity for two-component turbulence, the NADT model, and $\delta B / B_0 = 0.5$.
We have shown results for different values of the dissipation range spectral index $p$. The shaded band represents the Palmer (1982)
consensus range.}
\label{paracompdb05ppp}
\end{figure}

\begin{figure}
\centering
\includegraphics[scale=0.5]{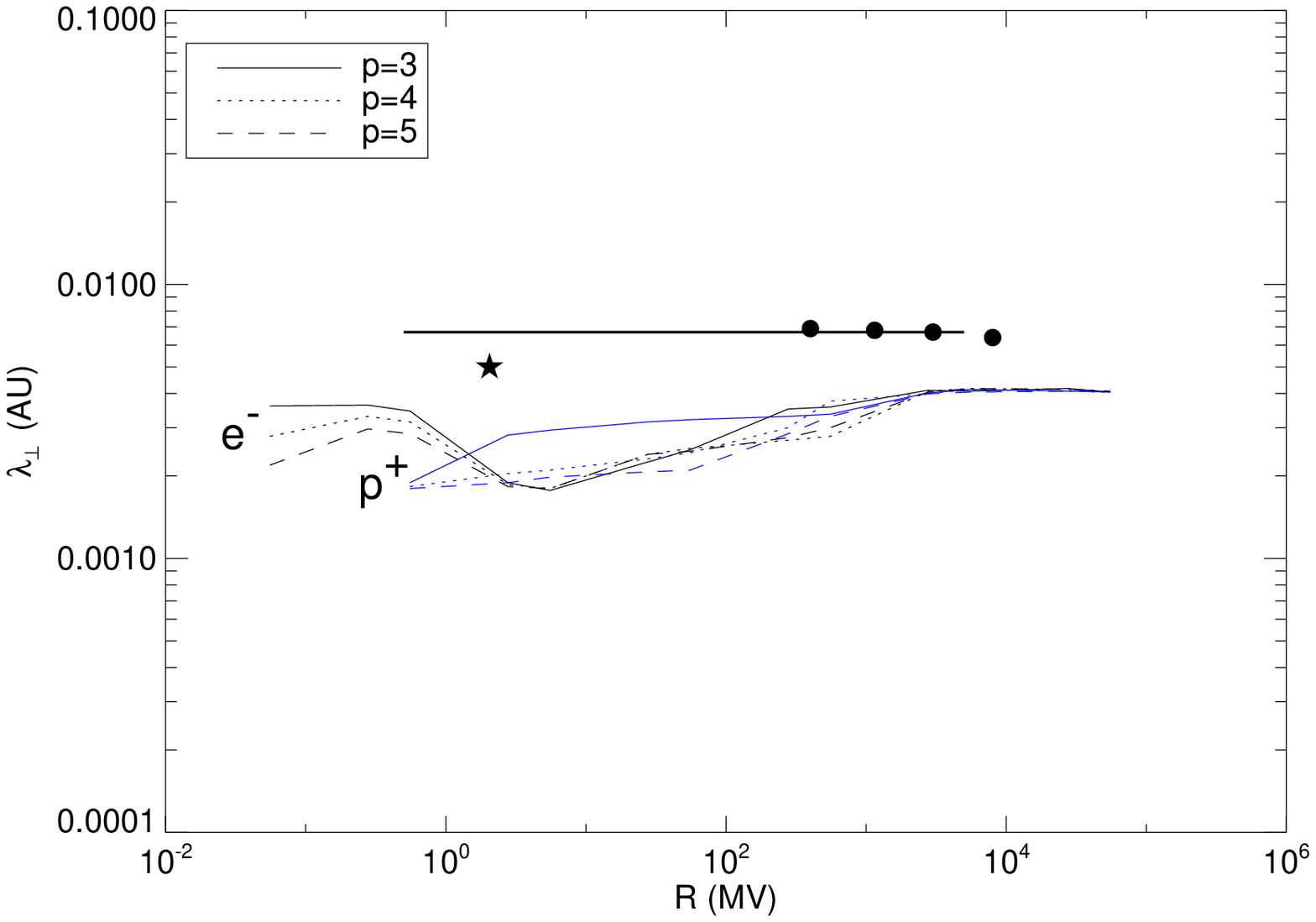}
\caption{The perpendicular mean free path versus magnetic rigidity for two-component turbulence, the NADT model, and $\delta B / B_0 = 0.5$.
We have shown results for different values of the dissipation range spectral index $p$. For comparison we show observations of Jovian
electrons (Chenette et al. 1977, star), Ulysses measurements of Galactic protons (Burger et al. 2000, dots), and the Palmer (1982) value
(horizontal line).}
\label{perpcompdb05ppp}
\end{figure}

\begin{figure}
\centering
\includegraphics[scale=0.5]{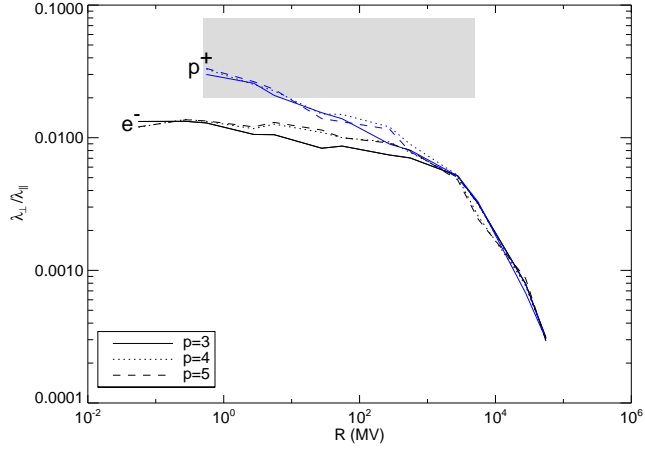}
\caption{The ratio of perpendicular and parallel mean free paths versus magnetic rigidity for two-component turbulence, the NADT model,
and $\delta B / B_0 = 0.5$. We have shown results for different values of the dissipation range spectral index $p$. The shaded band
represents the Palmer (1982) consensus range.}
\label{ratiocompdb05ppp}
\end{figure}

In Figs. \ref{paracompdb05ppp}, \ref{perpcompdb05ppp}, and \ref{ratiocompdb05ppp} we show the diffusion parameters for
$p=3$, $p=4$, and $p=5$. Obviously there is almost no influence of the dissipation range spectral index on the considered
transport parameters.

\subsection{Influence of the Two-dimensional Bendover Scale}
Another parameter which can be changed in our simulations, is the bendover scale of the two-dimensional modes $l_{2D}$. The latter parameter
denotes the turnover from the intermediate scales of the inertial range to the large scales of the energy range. Originally it was
assumed that $l_{2D} = 0.1 l_{slab}$, at least in the context of test-particle calculations (see again Bieber et al. 1994).
In recent years, the ratio of the two bendover scales was changed to $l_{2D} = l_{slab}$ (see, e.g., Hussein \& Shalchi 2016)
and this is what we have used above.

\begin{figure}
\centering
\includegraphics[scale=0.5]{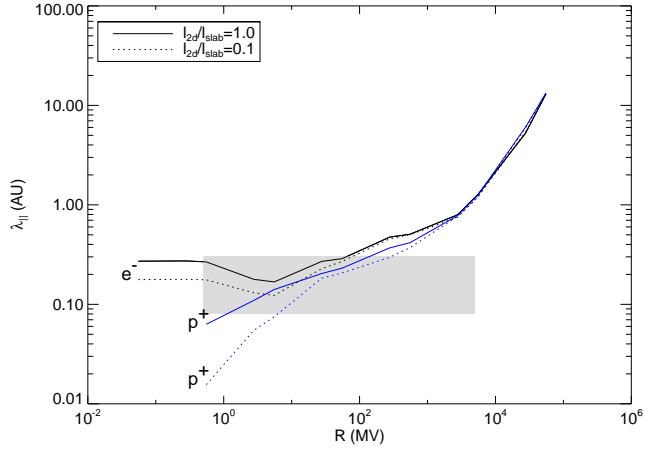}
\caption{The parallel mean free path versus magnetic rigidity for two-component turbulence, the NADT model, and
$\delta B / B_0 = 0.5$. We have shown results for different values of the two-dimensional bendover scale $l_{2D}$.
The shaded band represents the Palmer (1982) consensus range.}
\label{paracompdb05l2d}
\end{figure}

\begin{figure}
\centering
\includegraphics[scale=0.5]{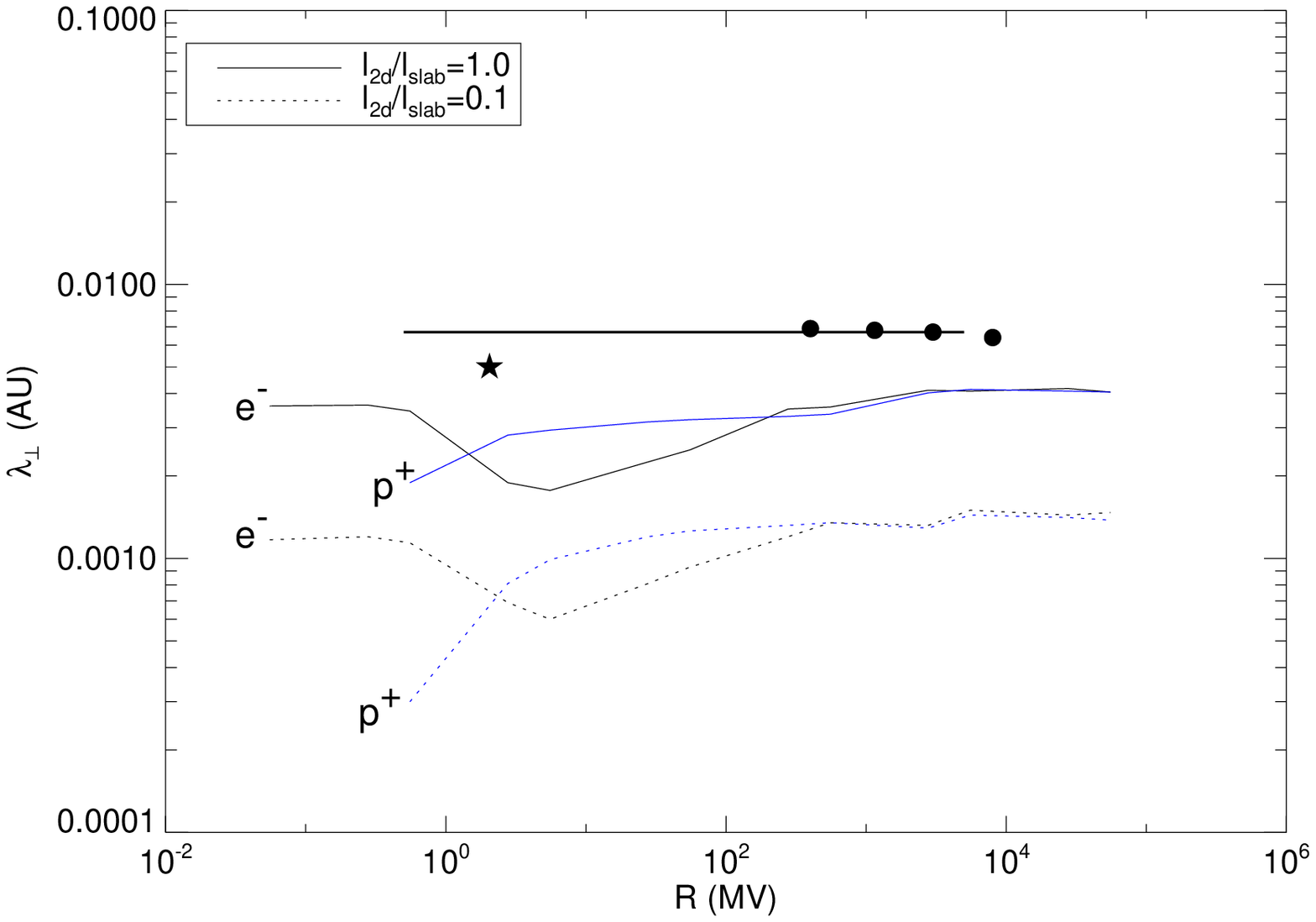}
\caption{The perpendicular mean free path versus magnetic rigidity for two-component turbulence, the NADT model,
and $\delta B / B_0 = 0.5$. We have shown results for different values of the two-dimensional bendover scale $l_{2D}$.
For comparison we show observations of Jovian electrons (Chenette et al. 1977, star), Ulysses measurements of Galactic
protons (Burger et al. 2000, dots), and the Palmer (1982) value (horizontal line).}
\label{perpcompdb05l2d}
\end{figure}

\begin{figure}
\centering
\includegraphics[scale=0.5]{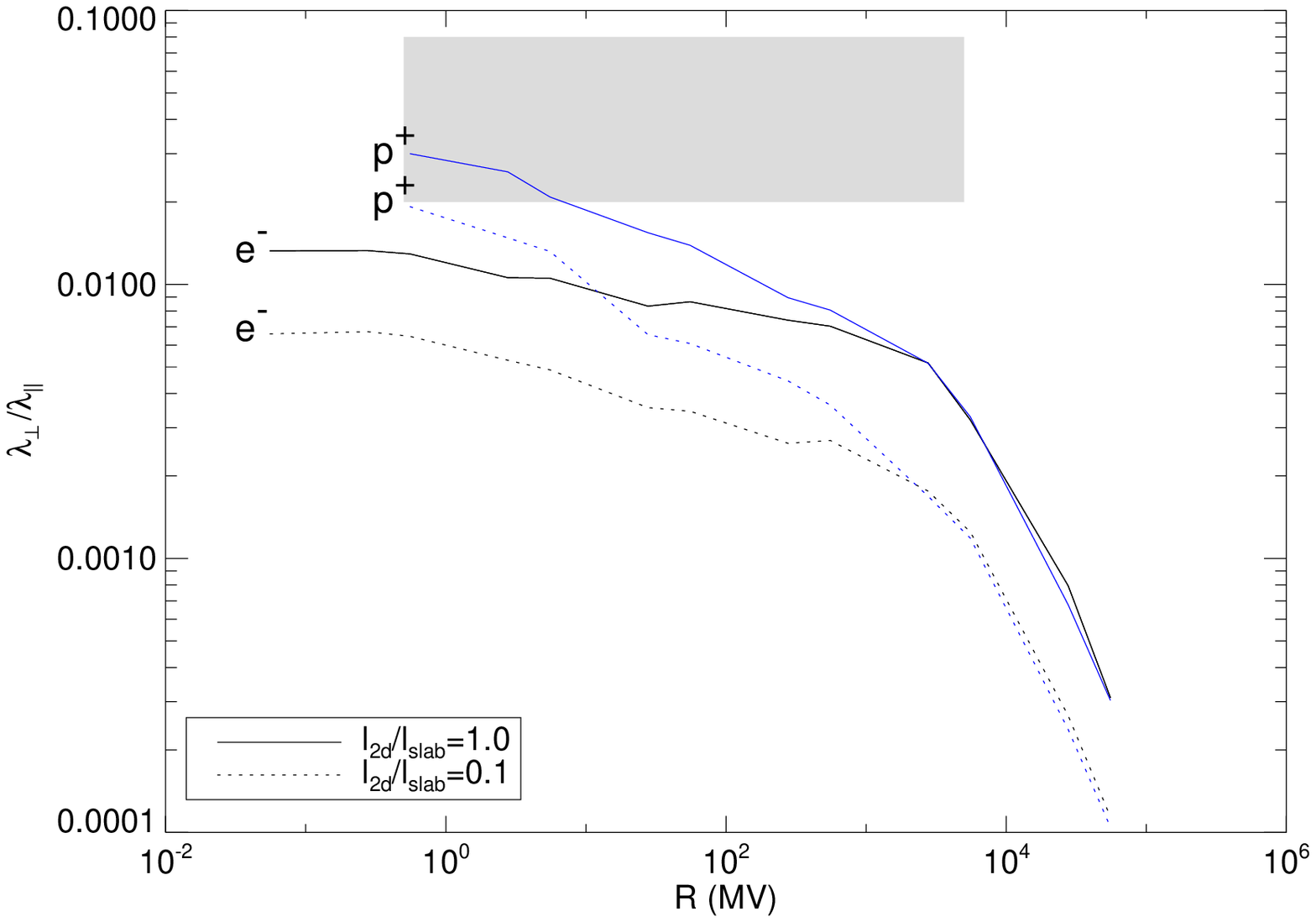}
\caption{The ratio of perpendicular and parallel mean free paths versus magnetic rigidity for two-component turbulence,
the NADT model, and $\delta B / B_0 = 0.5$. We have shown results for different values of the two-dimensional bendover
scale $l_{2D}$. The shaded band represents the Palmer (1982) consensus range.}
\label{ratiocompdb05l2d}
\end{figure}

In Figs. \ref{paracompdb05l2d}, \ref{perpcompdb05l2d}, and \ref{ratiocompdb05l2d} we show diffusion parameters for $l_{2D} = 0.1 l_{slab}$.
We can see that the parallel mean free path as well as the perpendicular mean free path are drastically reduced due to the smaller
values of $l_{2D}$. Clearly we find that the perpendicular mean free path is far away from the different interplanetary measurements.
The parallel mean free path, however, is now directly in the Palmer (1982) consensus range. The ratio $\lambda_{\perp} / \lambda_{\parallel}$
is too small as well.

\subsection{Influence of the Dissipation Scales}
Simulations of MHD turbulence in presence of a mean field display anisotropic power in the parallel and perpendicular direction.
In simulations the dissipative range is reached at different scales. The measure of the two-dimensional correlations and of the
Taylor scale in the solar wind (see Weygand et al. 2011) also support the existence of different dissipative scales in the parallel
and perpendicular directions. This corresponds to different dissipation wavenumbers $k_{d}^{2D}$ and $k_d^{slab}$.

\begin{figure}
\centering
\includegraphics[scale=0.5]{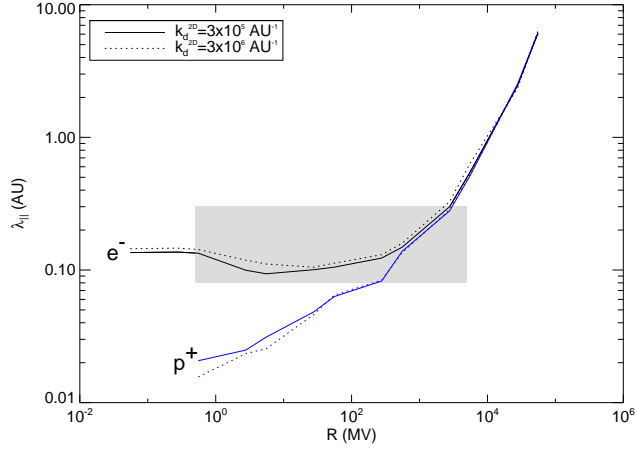}
\caption{The parallel mean free path versus magnetic rigidity for composite turbulence using the NADT model with $\delta B/B_0=0.75$,
$l_{2D}/l_{slab}=1.0$, and $p=3$ for different values of $k_d^{2D}$. The shaded band represents the Palmer (1982) consensus range.}
\label{prll_diss}
\end{figure}

\begin{figure}
\centering
\includegraphics[scale=0.5]{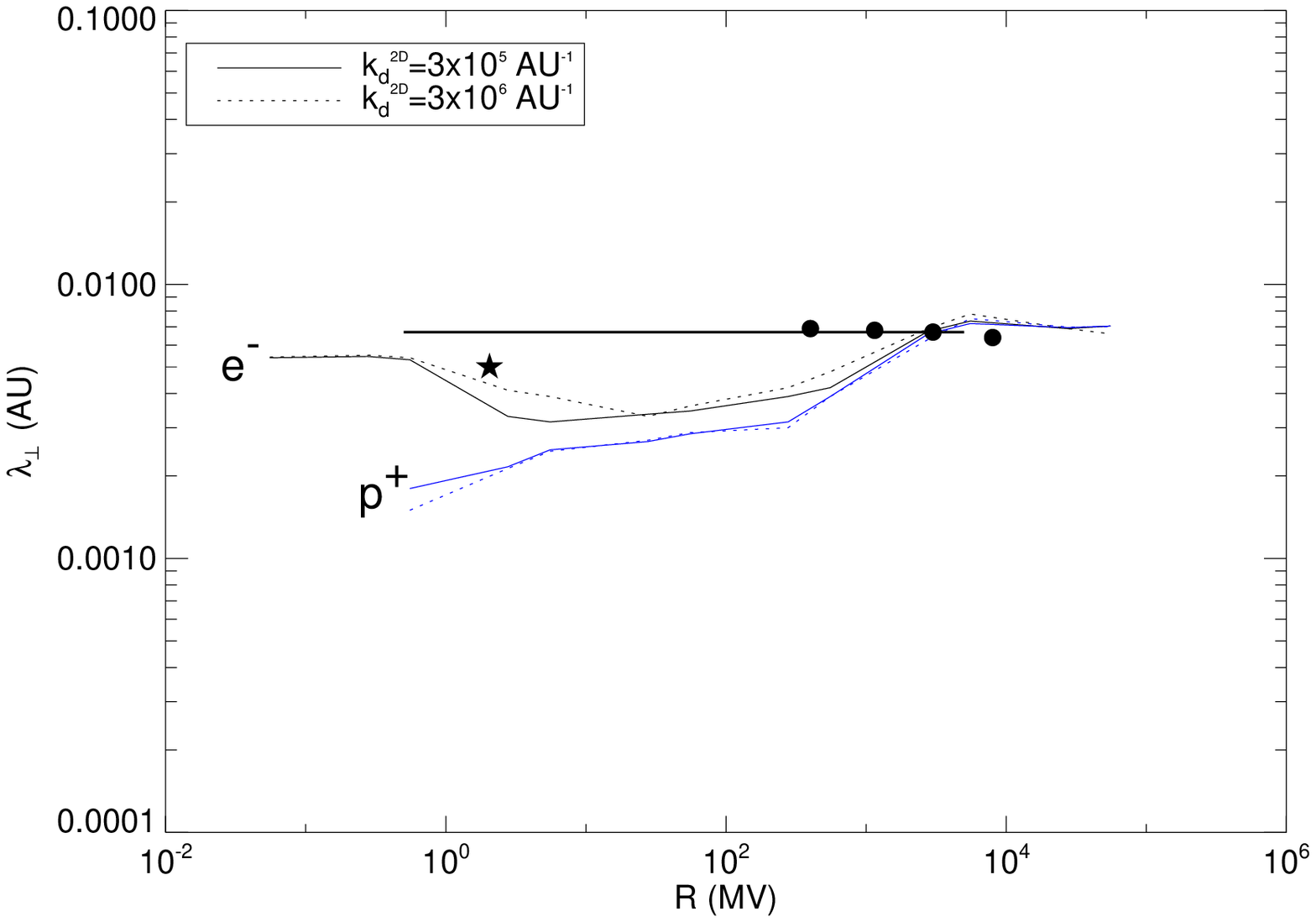}
\caption{The perpendicular mean free path versus magnetic rigidity for composite turbulence using the NADT model with $\delta B/B_0=0.75$,
$l_{2D}/l_{slab}=1.0$ and $p=3$ for different values of $k_d^{2D}$. We show observations of Jovian electrons (star), Ulysses measurements
of Galactic protons (dots), and the Palmer (1982) value (horizontal line).}
\label{perp_diss}
\end{figure}

\begin{figure}
\centering
\includegraphics[scale=0.5]{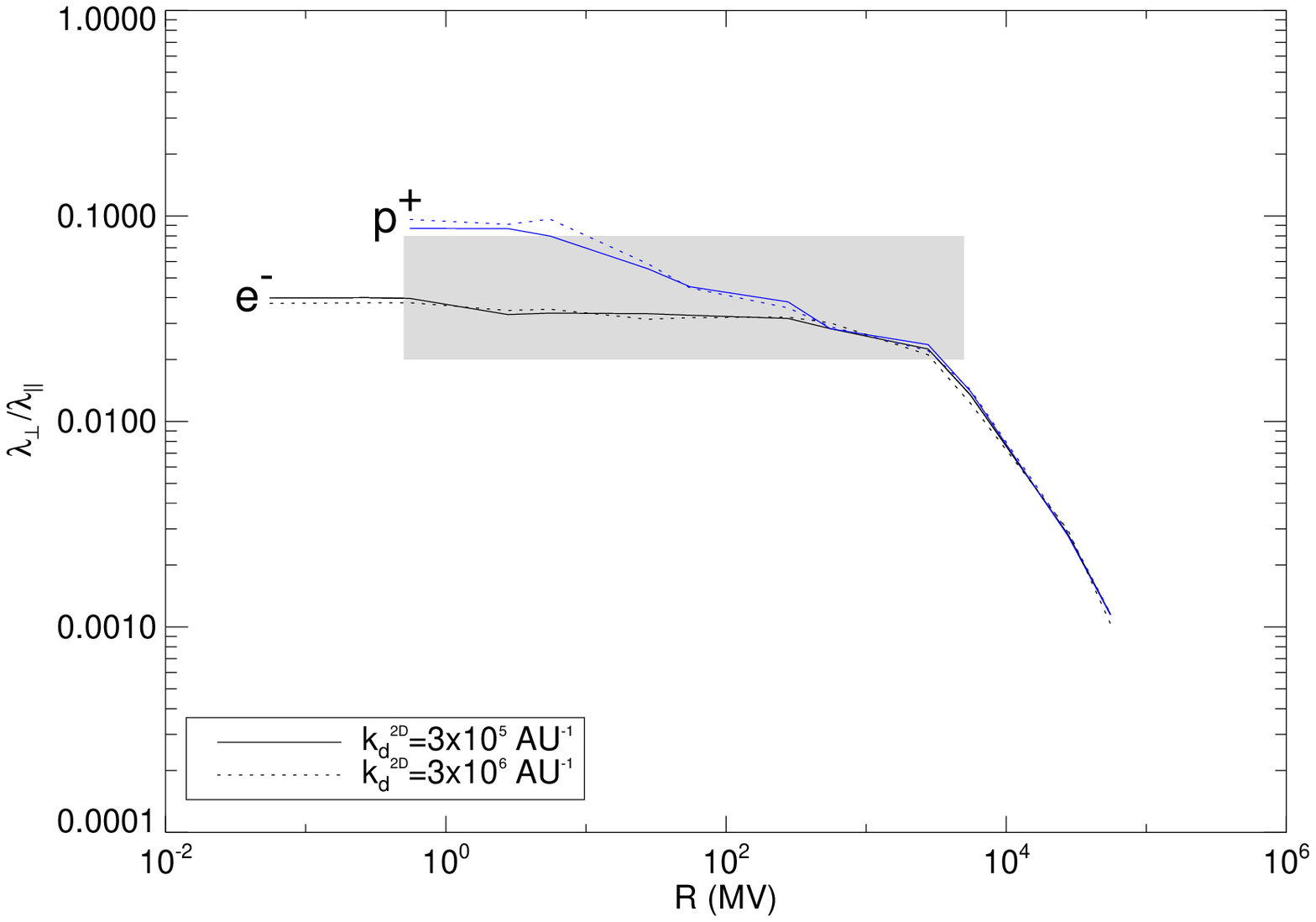}
\caption{The ratio of perpendicular to parallel mean free paths versus magnetic rigidity for composite turbulence using the NADT model
with $\delta B/B_0=0.75$, $l_{2D}/l_{slab}=1.0$, and $p=3$ for different values of $k_d^{2D}$. The shaded band represents the Palmer (1982)
consensus range.}
\label{perp_prll_diss}
\end{figure}

In order to test the influence of the dissipation scales on the diffusion of energetic particles, we repeat one set of simulations
with a higher value of the dissipation wavenumber of the two-dimensional modes $k_{d}^{2D}$. Above we have used $k_d=3 \times 10^5 1/AU$
in all of our simulations for both slab and the two-dimensional modes. We redo the set with $\delta B/B_0=0.75$, $l_{2D}=l_{slab}$, and $p=3$
keeping the slab dissipation wavenumber as is but use $k_d^{2D}=3 \times 10^6 1/AU$. Figs. \ref{prll_diss}, \ref{perp_diss}, and \ref{perp_prll_diss}
show the parallel mean free path, the perpendicular mean free path, and the ratio of the two mean free paths as function of rigidity
for the different values of $k_d^{2D}$. Clearly, the value of $k_d^{2D}$ has no noticeable influence on the transport parameters.

\subsection{Influence of the Inertial Range Spectral Index}
\begin{figure}
\centering
\includegraphics[scale=0.5]{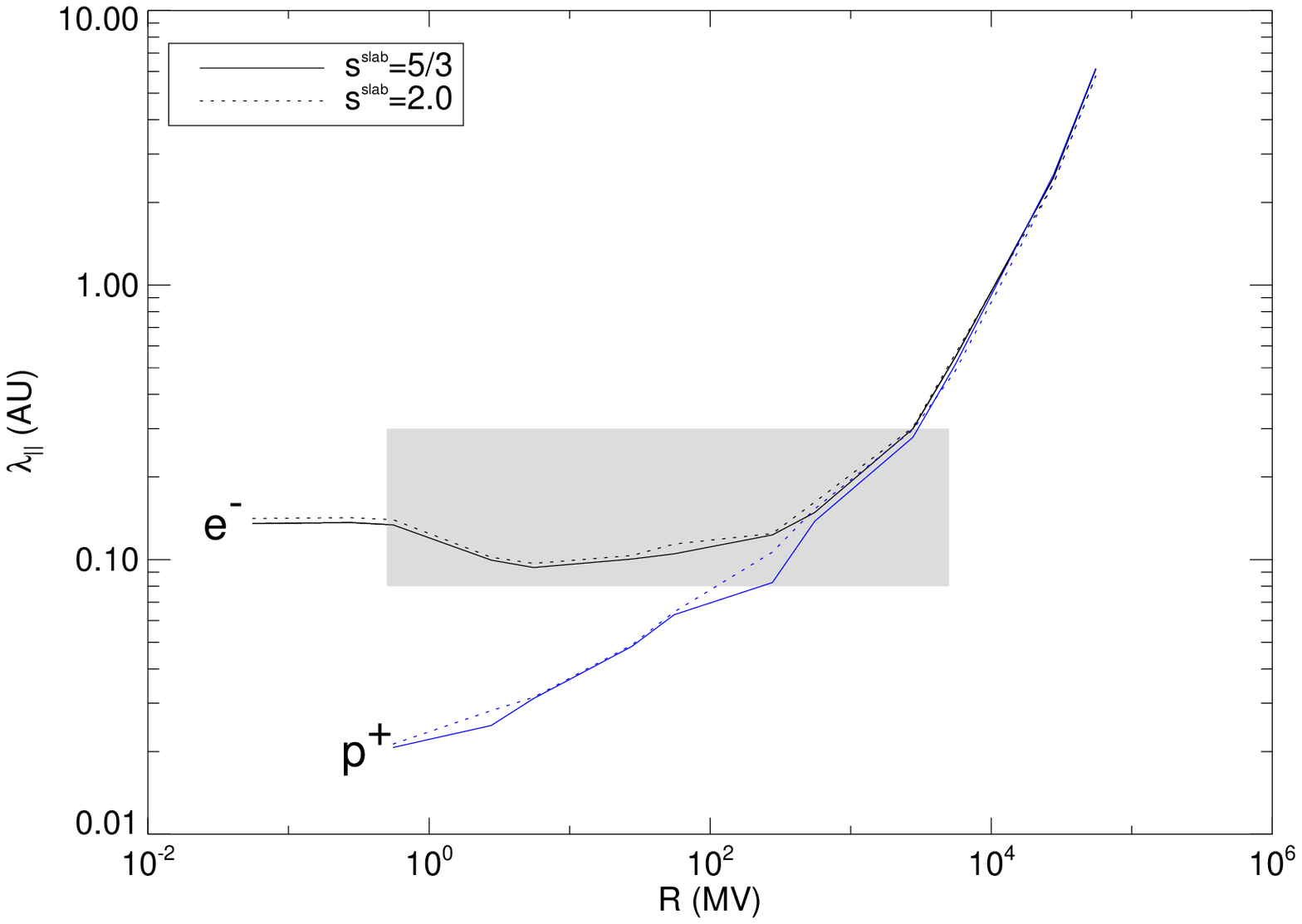}
\caption{The parallel mean free path versus magnetic rigidity for composite turbulence using the NADT model with $\delta B/B_0=0.75$,
$l_{2D}/l_{slab}=1.0$, and $p=3$ for different values of $s^{slab}$. The shaded band represents the Palmer (1982) consensus range.}
\label{prll_index}
\end{figure}

\begin{figure}
\centering
\includegraphics[scale=0.5]{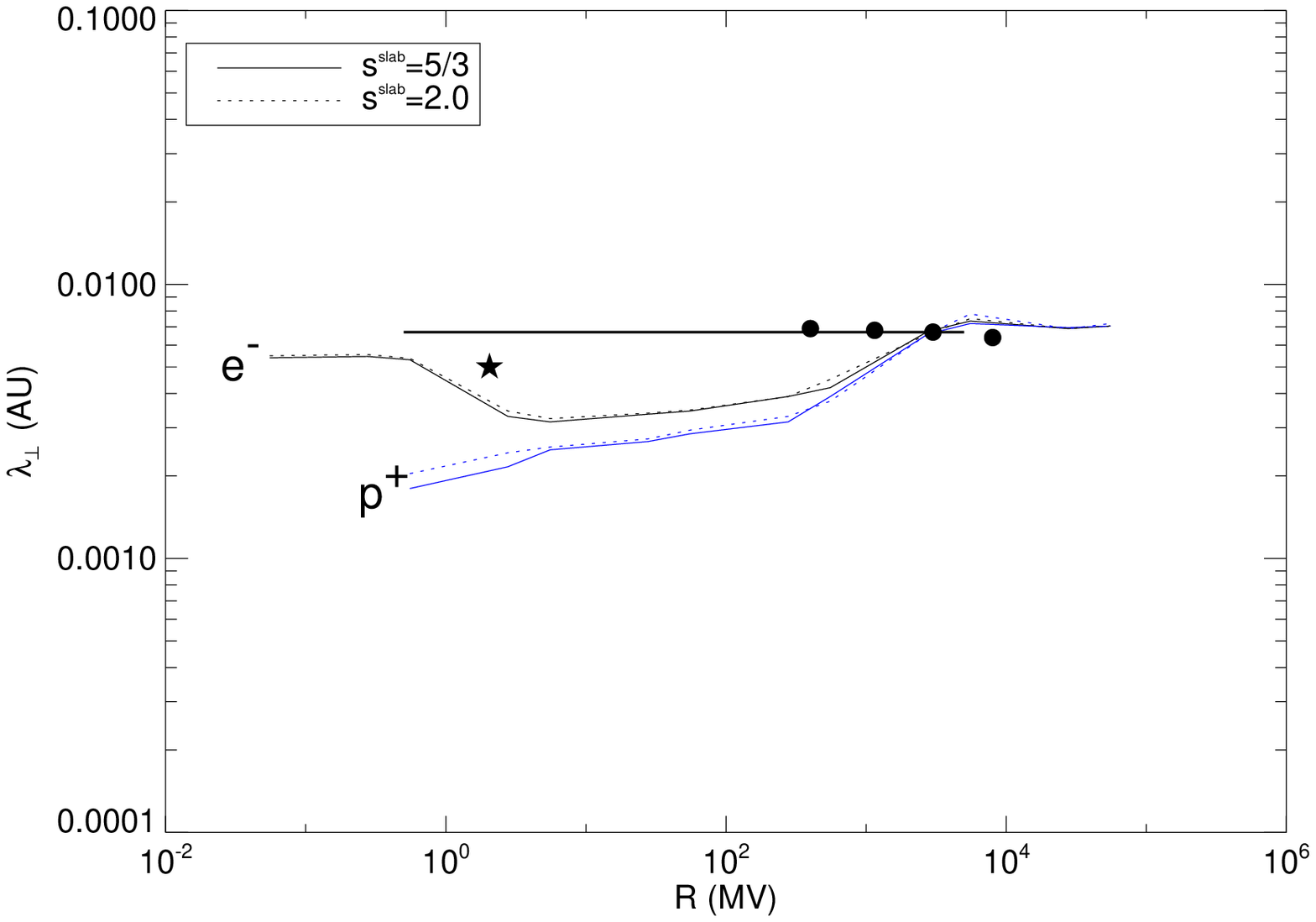}
\caption{The perpendicular mean free path versus magnetic rigidity for composite turbulence using the NADT model with $\delta B/B_0=0.75$,
$l_{2D}/l_{slab}=1.0$ and $p=3$ for different values of $s^{slab}$. We show observations of Jovian electrons (star), Ulysses measurements
of Galactic protons (dots), and Palmer (1982) value (horizontal line).}
\label{perp_index}
\end{figure}

\begin{figure}
\centering
\includegraphics[scale=0.5]{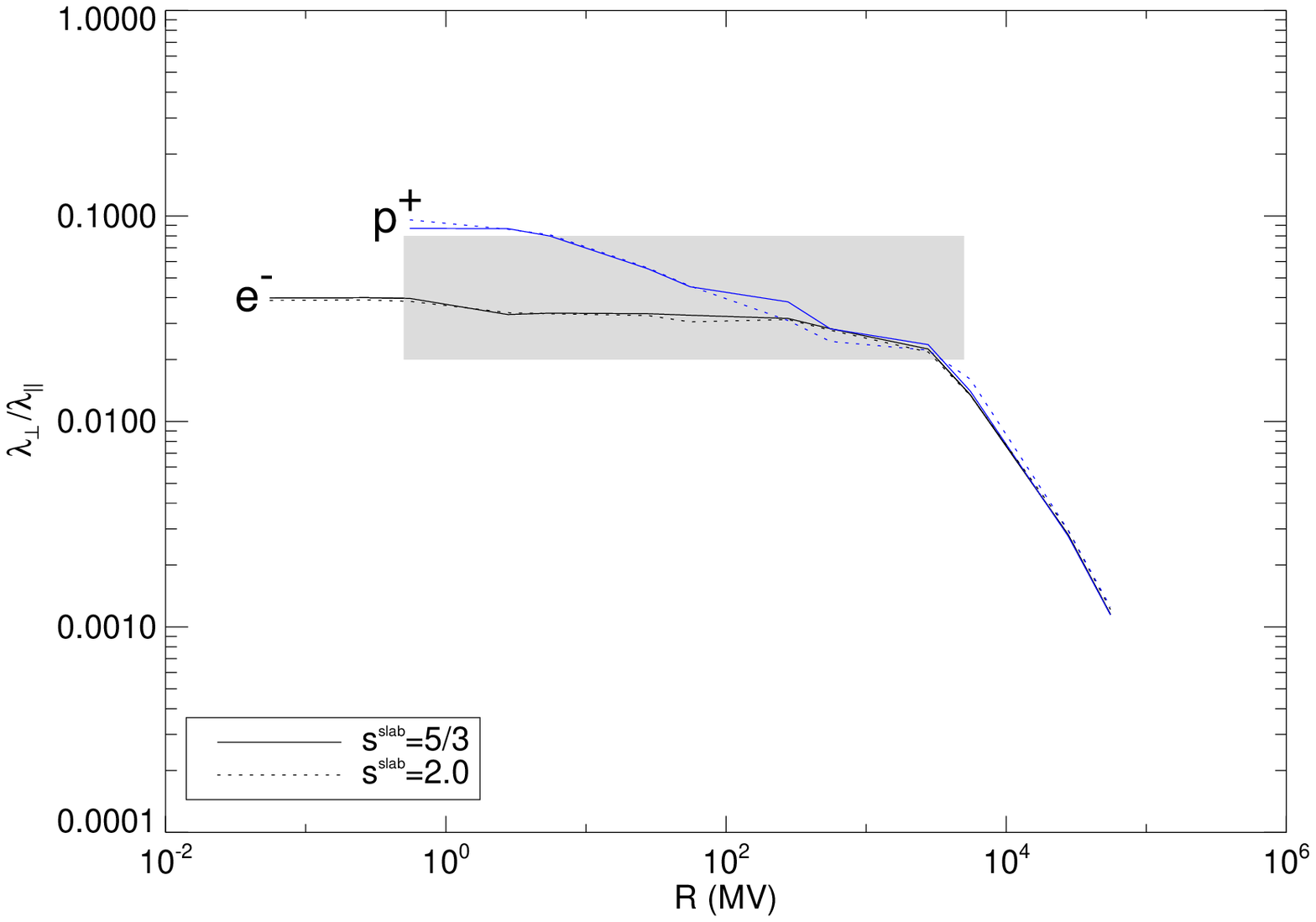}
\caption{The ratio of perpendicular to parallel mean free paths versus magnetic rigidity for composite turbulence using the NADT model
with $\delta B/B_0=0.75$, $l_{2D}/l_{slab}=1.0$, and $p=3$ for different values of $s^{slab}$. The shaded band represents the Palmer (1982)
consensus range.}
\label{perp_prll_index}
\end{figure}

In the local description of turbulence the parallel and perpendicular spectral indexes differ substantially (see, e.g., Goldreich \& Shridar 1995, Cho \& Vishniac 2000,
and Boldyrev 2005) and this has been confirmed by solar wind measurements (see, e.g., Horbury et al. 2008). To test the influence of a varying inertial range spectral index $s$
on the transport of energetic particles, we perform one set of simulations with $s^{slab}=2$ for the slab modes and keep $s^{2D}=5/3$ for the two-dimensional modes. As before, we use
$\delta B/B_0=0.75$, $l_{2D}=l_{slab}$, and $p=3$. Figs. \ref{prll_index}, \ref{perp_index}, and \ref{perp_prll_index} show the parallel mean free path, the perpendicular
mean free path, and the ratio of the two mean free paths as function of rigidity for the different values of $s^{slab}$. Clearly, a steeper inertial range for the slab modes
has no noticeable influence on the transport parameters.

\section{Summary and Conclusion}
The current paper is a sequel of Hussein \& Shalchi (2016) where we have started to perform test-particle simulations
for dynamical turbulence. The turbulence dynamics can have a strong influence on particle diffusion coefficients at
low particle rigidities. In the previous work, we have employed two models for dynamical turbulence, namely the
{\it damping model of dynamical turbulence} and the {\it random sweeping model}. Both models were originally proposed
in the pioneering work of Bieber et al. (1994).

It is the purpose of the current paper to replace the aforementioned dynamical turbulence models by the so-called
{\it Nonlinear Anisotropic Dynamical Turbulence (NADT) model} of Shalchi et al. (2006) which takes into account wave
propagation effects as well as damping effects. Furthermore, we perform a detailed parameter study in order to explore
the influence of the magnetic field ratio $\delta B / B_0$, the turbulence scale ratio $l_{2D} / l_{slab}$, the dissipation
range spectral index $p$, the dissipation wavenumber $k_d$, as well as the inertial range spectral index $s$ on the parallel
mean free path $\lambda_{\parallel}$, the perpendicular mean free path $\lambda_{\perp}$, and the ratio of the two diffusion
parameters $\lambda_{\perp}/\lambda_{\parallel}$. Our findings are shown in Figs. \ref{paracompdb1}-\ref{perp_prll_index}
and the corresponding parameter values are listed in Tables \ref{simvalues} and \ref{runs}.

We found that the influence of the dissipation range spectral index is minor. The influence of the inertial range
spectral index and the dissipation scales are negligible as well. The magnetic field ratio, on the other hand,
has a strong influence on both diffusion coefficients and their ratio. We found best agreement with the Palmer (1982)
consensus range for $\delta B / B_0 = 0.75$ (corresponding to approximately $\delta B^2 / B_0^2 = 0.6$) which is between
the values $\delta B / B_0 = 0.5$ and $\delta B / B_0 = 1$ usually used for this type of work. We also found that the ratio
of the bendover scales $l_{2D} / l_{slab}$ has an influence on the parallel mean free path and a very strong influence
on the perpendicular diffusion coefficient. This was predictable because analytical treatments of the transport
(see, e.g., Shalchi 2015) show the importance of the so-called {\it Kubo number} on the perpendicular motion of energetic
particles. The latter number depends on the magnetic field ratio as well as the turbulence scales.

The main conclusion of the current paper is that we can indeed reproduce different solar wind observations performed for
energetic particles interacting with magnetic turbulence if we employ the NADT model. More detailed turbulence measurements
would show what the exact value of the different parameters used in the current paper are. Then one could draw more conclusions
concerning the validity of the employed turbulence model.

\acknowledgements
{\it A. Shalchi acknowledges support by the Natural Sciences and Engineering Research Council (NSERC) of Canada.
Most simulations shown in this article were obtained by using the national computational facility provided by WestGrid.
We are also grateful to S. Safi-Harb for providing her CFI-funded computational facilities for code tests and for some of
the simulation runs presented here.}
{}


\begin{thebibliography}{}

\bibitem[Alexandrova et al.(2008)]{Alexandrova08}
Alexandrova, O., Carbone, V., Veltri, P., \& Sorriso-Valvo, L. 2008, ApJ, 674, 1153

\bibitem[Bavassano(2003)]{bav03}
Bavassano, B. 2003, AIP Conference Proceedings, 679, 377

\bibitem[Belcher \& Davis(1971)]{belcher71}
Belcher, J. W., \& Davis Jr., L. 1971, JGR, 76, 3534

\bibitem[Bieber et al.(1994)]{bie94}
Bieber, J. W., Matthaeus, W. H., Smith, C. W., Wanner, W., Kallenrode, M.-B., \& Wibberenz, G. 1994, ApJ, 420, 294

\bibitem[Bieber et al.(1996)]{bieber96}
Bieber, J. W., Wanner, W., \& Matthaeus, W. H. 1996, JGR, 101, 2511

\bibitem[Boldyrev(2005)]{Boldyrev2005}
Boldyrev, S. 2005, \apjl, 626, L37

\bibitem[Burger et al.(2000)]{burger00}
Burger, R. A., Potgieter, M. S., \& Heber, B. 2000, JGR, 105, 27447

\bibitem[Chenette et al.(1977)]{chenette77}
Chenette, D. L., Conlon, T. F., Pyle, K. R., \& Simpson, J. A. 1977, ApJ, 215, L95

\bibitem[Cho \& Vishniac(2000)]{Cho2000}
Cho, J. \& Vishniac, E.~T. 2000, \apj, 539, 273

\bibitem[Denskat \& Neubauer(1982)]{Denskat82}
Denskat, K. U. \& Neubauer, F. M. 1982,
Observations of hydrodynamik turbulence in the solar wind.
{\it In Solar Wind Five, (Ed.) Neugebauer, M., Proceedings of a conference
held in Woodstock, Vermont, November 1-5, 1982, vol. 2280 of NASA
Conference Publication, pp. 81-91, NASA, Washington, U.S.A}

\bibitem[Dong et al.(2014)]{Dong14}
Dong, Y., Verdini, A., \& Grappin, R. 2014, ApJ, 793, 118

\bibitem[Giacalone \& Jokipii(1999)]{giacalone99}
Giacalone, J., \& Jokipii, J. R. 1999, ApJ, 520, 204

\bibitem[Goldreich \& Sridhar(1995)]{Goldreich95}
Goldreich, P. \& Sridhar, S. 1995, \apj, 438, 763

\bibitem[Horbury et al.(2008)]{Horbury08}
Horbury, T.~S., Forman, M., \& Oughton, S. 2008, PRL, 101, 175005

\bibitem[Howes et al.(2008)]{Howes08}
Howes, G. G., Dorland, W., Cowley, S. C., Hammett, G. W., Quataert, E., Schekochihin, A. A., \& Tatsuno, T. 2008, 100, 065004

\bibitem[Hussein et al.(2014)]{huss2014}
Hussein, M. \& Shalchi, A. 2014, ApJ, 785, 31

\bibitem[Hussein et al.(2015)]{huss2015}
Hussein, M., Tautz, R., \& Shalchi, A. 2015, JGR, 120, 4095

\bibitem[Hussein \& Shalchi(2016)]{huss2016}
Hussein, M. \& Shalchi, A. 2016, ApJ, 817, 136

\bibitem[King(1989)]{King89}
King, J. H. 1989, Interplanetary medium data book, supplement 4, 1985-1988

\bibitem[Matthaeus et al.(1990)]{matt90}
Matthaeus, W. H., Goldstein, M. L., \& Roberts, D. A. 1990, JGR, 95, 20673

\bibitem[Matthaeus et al.(1996)]{matt96}
Matthaeus, W. H., Ghosh, S., Oughton, S., \& Roberts, D. 1996, JGR, 101, 7619

\bibitem[Matthaeus et al.(2007)]{matt07}
Matthaeus, W. H., Bieber, J. W., Ruffolo, D., Chuychai, P., \& Minnie, J. 2007, ApJ, 667, 956

\bibitem[Micha\l ek \& Ostrowski(1996)]{mic96}
Micha\l ek, G., \& Ostrowski, M. 1996, Nonlin. Proc. Geophys., 3, 66

\bibitem[Narita et al.(2010)]{Narita10}
Narita, Y., Glassmeier, K.-H., Sahraoui, F., \& Goldstein, M. L. 2010, PRL, 104, 171101

\bibitem[Osman \& Horbury(2009a)]{osmhor09a}
Osman, K. T., \& Horbury, T. S. 2009a, JGR, 114, A06103

\bibitem[Osman \& Horbury(2009b)]{osmhor09b}
Osman, K. T., \& Horbury, T. S. 2009b, Annales Geophysicae, 27, 3019

\bibitem[Oughton et al.(1994)]{ought94}
Oughton, S., Priest, E. R., \& Matthaeus, W. H. 1994, J. Fluid. Mech., 280, 95

\bibitem[Oughton et al.(2006)]{ot06}
Oughton, S., Dmitruk, P., \& Matthaeus, W. H. 2006, Phys. Plasmas, 13, 042306

\bibitem[Palmer(1982)]{palmer82}
Palmer, I. D. 1982, Rev. Geophys. Space Phys., 20, 335

\bibitem[Qin et al.(2002a)]{qin2002a}
Qin, G., Matthaeus, W. H., \& Bieber, J. W. 2002a, GeoRL, 29, 1048

\bibitem[Qin et al.(2002b)]{qin02b}
Qin, G., Matthaeus, W. H., \& Bieber, J. W., 2002b, ApJ, 578, L117

\bibitem[Ruffolo et al.(2012)]{ruffolo2012}
Ruffolo, D.,  Pianpanit, T., Matthaeus, W. H., \& Chuychai, P. 2012, ApJ, 747, L34

\bibitem[Saur \& Bieber(1999)]{Saur99}
Saur, J. \& Bieber, J. W. 1999, JGR, 104, 9975

\bibitem[Schlickeiser(2002)]{Schlick02}
Schlickeiser, R., 2002, Cosmic Ray Astrophysics (Berlin: Springer)

\bibitem[Shaikh \& Zank(2007)]{shaik07}
Shaikh, D., \& Zank, G. P. 2007, ApJ, 656, L17

\bibitem[Shalchi et al.(2006)]{shal06}
Shalchi, A., Bieber, J. W., Matthaeus, W. H., \& Schlickeiser, R. 2006, ApJ, 642, 230

\bibitem[Shalchi(2009)]{shal09book}
Shalchi, A. 2009, Nonlinear Cosmic Ray Diffusion Theories (Astrophysics and Space Science Library, Vol. 362; Berlin: Springer)

\bibitem[Shalchi \& Weinhorst(2009)]{Shalwei2009}
Shalchi, A., \& Weinhorst, B. 2009, AdSpR, 43, 1429

\bibitem[Shalchi(2015)]{shal2015}
Shalchi, A. 2015, PhPl, 22, 010704

\bibitem[Tautz(2010)]{tautz10}
Tautz, R. C. 2010, Comput. Phys. Commun., 181, 71

\bibitem[Tautz \& Shalchi(2013)]{taushalpalmer13}
Tautz, R. C., \& Shalchi, A. 2013, JGR, 118, 642

\bibitem[Tu \& Marsch(1993)]{Tu93}
Tu, C.-Y. \& Marsch, E. 1993, JGR, 98, 1257

\bibitem[Turner et al.(2012)]{turn12}
Turner, A. J., Gogoberidze, G., \& Chapman. S. C. 2012, PhRvL, 108, 8

\bibitem[Weinhorst et al.(2008)]{Weinhorst08}
Weinhorst, B., Shalchi, A., \& Fichtner, H. 2008, ApJ, 677, 671

\bibitem[Weygand et al.(2011)]{Weygand2011}
Weygand, J.~M., Matthaeus, W.~H., Dasso, S., \& Kivelson, M.~G.\ 2011, Journal of Geophysical Research (Space Physics), 116, A08102

\bibitem[Zank \& Matthaeus(1993)]{zankmat93}
Zank, G. P., \& Matthaeus, W. H. 1993, Physics of Fluids A, 5, 257

\bibitem[Zhou et al.(2004)]{zhou04}
Zhou, Y., Matthaeus, W. H., \& Dmitruk, P. 2004, Rev. Mod. Phys., 76, 1015

\end{thebibliography}
\end{document}